\documentclass[parskip=full,11pt]{article}
\usepackage{amssymb}
\usepackage{amsthm}
\usepackage{amsmath}
\usepackage{float,booktabs}
\usepackage{float} 
\usepackage{graphicx}
\usepackage{natbib}
\usepackage{url}
\usepackage{xcolor}
\usepackage{soul,color,colortbl}
\usepackage{array,multirow}
\usepackage{cancel}
\usepackage[top=0.9in, bottom=1.2in, left=0.875in, right=0.875in]{geometry}
\usepackage[linesnumbered,ruled]{algorithm2e}
\usepackage{tikz}
\usetikzlibrary{er,positioning,arrows.meta,calc}
\usepackage[labelformat=simple]{subcaption}
\usepackage{bm}
\usepackage[linesnumbered,ruled]{algorithm2e}
\usepackage{caption}
\usepackage{subcaption}
\usepackage[linesnumbered,ruled]{algorithm2e}
\usetikzlibrary{shapes,arrows,shadows}
\usepackage{bm}
\usepackage{bbm}
\usepackage[titletoc,toc,title]{appendix}
\usepackage{tablefootnote}

\newcolumntype{L}[1]{>{\raggedright\let\newline\\\arraybackslash\hspace{0pt}}m{#1}}
\newcolumntype{C}[1]{>{\centering\let\newline\\\arraybackslash\hspace{0pt}}m{#1}}
\newcolumntype{R}[1]{>{\raggedleft\let\newline\\\arraybackslash\hspace{0pt}}m{#1}}

\def\spacingset#1{\renewcommand{\baselinestretch}
{#1}\small\normalsize} \spacingset{1}

\newcolumntype{P}[1]{>{\centering\arraybackslash}p{#1}}
\newcommand*{\myfont}{\fontfamily{lmss}\selectfont}
\DeclareTextFontCommand{\textpython}{\myfont}

\DeclareCaptionLabelFormat{boldnumber}{\bothIfFirst{#1}{~}\textbf{#2}}
\title{\textbf{Improving Business Insurance Loss Models by Leveraging InsurTech Innovation}}

\author{Zhiyu Quan \thanks{Corresponding author; Department of Mathematics, University of Illinois, 1409 W. Green Street, Urbana, IL, 61801, USA. Email: \texttt{zquan@illinois.edu}.}
\and Changyue Hu\thanks{Department of Mathematics, University of Illinois, 1409 W. Green Street, Urbana, IL, 61801, USA. Email: \texttt{ch47@illinois.edu}.}
\and Panyi Dong\thanks{Department of Mathematics, University of Illinois, 1409 W. Green Street, Urbana, IL, 61801, USA. Email: \texttt{panyid2@illinois.edu}.}
\and Emiliano A. Valdez\thanks{Department of Mathematics, University of Connecticut, 341 Mansfield Road, Storrs, CT, 06269-1009, USA. Email: \texttt{emiliano.valdez@uconn.edu}.}}

\begin{document}

\maketitle

\begin{abstract}

Recent transformative and disruptive advancements in the insurance industry have embraced various InsurTech innovations. In particular, with the rapid progress in data science and computational capabilities, InsurTech is able to integrate a multitude of emerging data sources, shedding light on opportunities to enhance risk classification and claims management. This paper presents a groundbreaking effort as we combine real-life proprietary insurance claims information together with InsurTech data to enhance the loss model, a fundamental component of insurance companies' risk management. Our study further utilizes various machine learning techniques to quantify the predictive improvement of the InsurTech-enhanced loss model over that of the insurance in-house. The quantification process provides a deeper understanding of the value of the InsurTech innovation and advocates potential risk factors that are unexplored in traditional insurance loss modeling. This study represents a successful undertaking of an academic-industry collaboration, suggesting an inspiring path for future partnerships between industry and academic institutions.

\vspace{1.0cm}


\end{abstract}

\newpage

\section{Introduction}

\begin{figure}[h!]
  \centering
    \includegraphics[width=0.9\linewidth]{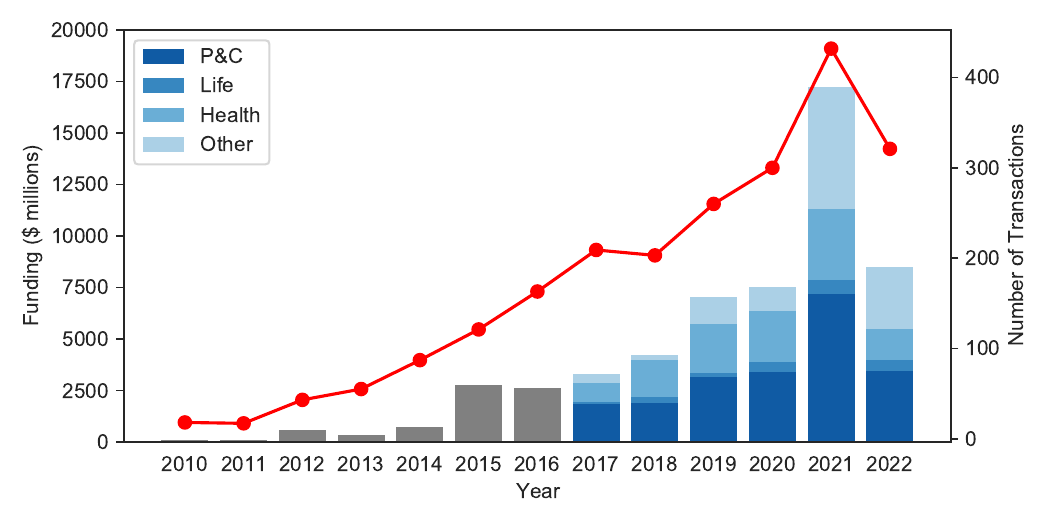}
    \caption{InsurTech financing for 2010-2022\protect\footnotemark.}
  \label{fig:financing}
\end{figure}

InsurTech combines the words ``insurance" and ``technology" to describe the applications of emerging technology to modernize the entire insurance value chain by improving efficiency, enriching customer service, enhancing underwriting and actuarial processes, and further uncovering new opportunities, to name a few. InsurTech emerged around 2010 and was initially categorized as a subsector of FinTech \citep{kelley2021}. InsurTech companies are conventionally funded as start-up companies by venture capitalists; InsurTech reached more than 2,000 start-up deals at the end of 2022. The amount of financing has been on the rise globally over the past decade, and in 2021, funds invested total 17.2 billion, breaking records of amounts invested in 2019 and 2020 combined. See Figure \ref{fig:financing}. The primary investment is directed towards Property and Casualty (P\&C)-focused InsurTech, followed by Health, with the lowest allocation to Life. While investment dropped in 2022 due to disruption in supply chains and economic uncertainties affecting inflation and interest rates, InsurTech financing will continue to have a significant presence in the industry for years ahead, with huge potential growth opportunities in the Life sector. We continue to see increasing evidence that InsurTech has a significant impact on the local insurance industry in places such as Europe \citep{ricci2021}, China \citep{Wang2021}, India \citep{suryavanshi2022}, and Israel \citep{berman2021business}.

\footnotetext{Data sourced from Financial Technology Partners' InsurTech Insights Reports. Retrieved from: \url{https://www.ftpartners.com/fintech-research/insurtech-almanac}}

The progress in InsurTech is accelerated by the rapid increase in the development and adoption of technology, much of which results in improved connectivity and modern digitization \citep{fluckiger2022}. Such technological advancements lead to the flow of big data that allows InsurTech innovation to offer user-oriented or data-driven solutions across all major lines of business in the insurance industry, including, as alluded earlier, the automation of business processes, the development of innovative products, and the exploitation of data for underwriting, risk assessment, and claims handling. See \cite{vanderlinden2018InsurTech}, \cite{xu2020framework}.

Two of the representative InsurTech applications are telematics in auto insurance and wearable technologies in the life and health sector. For more than a decade, auto insurance companies have used telematics technology to collect information on driving behavior and conditions that would otherwise be unavailable through conventional underwriting \citep{guillen2019use}. Such information includes, for example, exposure to traffic conditions, driving maneuvering, and distance traveled in real time, which are then used to improve risk assessment, explore the potential impact on pricing, and determine more suitable premiums for usage-based insurance. See \cite{boucher2017exposure}, \cite{weidner2017telematic}, \cite{verbelen2018unravelling}, \cite{gao2019claims}, \cite{huang2019automobile}, and \cite{che2021usage}. Companies like Discovery Insure, through their Vitality Drive program\footnote{\url{https://www.discovery.co.za/portal/individual/insure-vitality-drive}}, exemplify the application of telematics data, rewarding safe driving behavior with lower premiums, fuel cash back, and real-time emergency aid. Root Insurance\footnote{\url{https://www.joinroot.com}}, the nation’s first licensed insurance carrier powered entirely by telematics, offers a mobile application to monitor actual driving behavior, allowing a more customized pricing that is suitably affordable for safer drivers, but unattractive for reckless drivers. Studies on Root Insurance's financial reports highlight a noteworthy high combined ratio, prompting further research into the impact of telematics data on insurance claims. \cite{pesantez2019predicting} provide extensive numerical analysis on how telematics data can help understand the claim occurrences. In the cases where the telematics data cannot be directly associated with insurance claims, \cite{guillen2020can} propose to introduce near-miss events, which include potentially dangerous behaviors, based on certain manually-defined criteria into telematics data to identify risks. The near-miss telematics, suggested by \cite{guillen2021}, can be utilized as extra ratemaking factors for premium adjustments and incentives to promote safe driving. Furthermore, \cite{pesantez2019predicting}, \cite{weidner2017telematic} and \cite{gao2019claims} emphasize the importance of interpretability of InsurTech variables in the insurance industry and illustrate their models through model coefficients or feature importance. We also witness comparable technological advancements propelled by InsurTech in the healthcare sector. Wearable technology, or ``wearables,'' is growing in popularity and is used in the life and health insurance industry to engage customers, collect nontraditional data, and help with underwriting, pricing, loss mitigation, and customer retention \citep{spender2019wearables}.  For example, health insurance companies partner with Vitality Health International\footnote{\url{http://www.vitalityhealthinternational.com}}, empowered by the world's largest behavioral engagement platform, to track health measurements and reward healthy behavior for customers seeking a healthier lifestyle. \cite{mccrea2018wearable} develop a health risk score to investigate how wearable technology could be used to predict overall health and mortality, and to improve the pricing of health insurance. 

The InsurTech innovations discussed earlier harness emerging technology to provide additional information, empowering insurers to revolutionize conventional insurance models for strategic advantages, but fail to provide empirical evidence on how insurance pricing can be influenced. In this paper, we aim to fill this gap by \textit{quantifying} the impact of leveraging InsurTech data to improve loss models, which can then be utilized for ratemaking and risk classification. Consequently, insurers can gain a better understanding of the risks within their insurance portfolios, handle claims more efficiently, and further promote more effective marketing. To be more precise, we use data from an InsurTech start-up, Carpe Data\footnote{\url{https://www.carpe.io}}, to examine the efficacy of InsurTech in improving loss models for the entire portfolio of Business Owner's Policy (BOP) insurance from a mid-sized multi-state insurance company. After real-life data analysis and actuarial loss modeling, we are able to demonstrate that InsurTech-enhanced models exhibit improved predictability with viable interpretability compared with the existing insurance in-house model. This improvement is attributed to the additional information collected by Carpe Data.

To the best of our knowledge, there are no published articles in the actuarial literature on InsurTech innovation in business insurance, also known as commercial insurance, for a few reasons. First, \cite{chester2018commercial} point out that the complexity of underwriting and claims, combined with the low volume and the customized nature of transactions, poses obstacles to business insurance that embraces InsurTech. Second, to take advantage of the comprehensive capacity and fully unlock the potential benefits of InsurTech, combining and centralizing data from both insurers and InsurTech is critical, which may pose another challenge for researchers; even in practice, this kind of data sharing is a constraint due to privacy issues. However, we cannot undermine the importance of business insurance. Business insurance premiums are approximately equal to 2.3\% of the world's GDP\footnote{Data extracted from: Organisation for Economic Co-operation and Development (OECD) \url{https://stats.oecd.org/Index.aspx?DatasetCode=SNA_TABLE1}}, an indispensable share which makes the study of business insurance pressing at present.

\begin{figure}[h!]
  \centering
    \includegraphics[width= 1.0\linewidth]{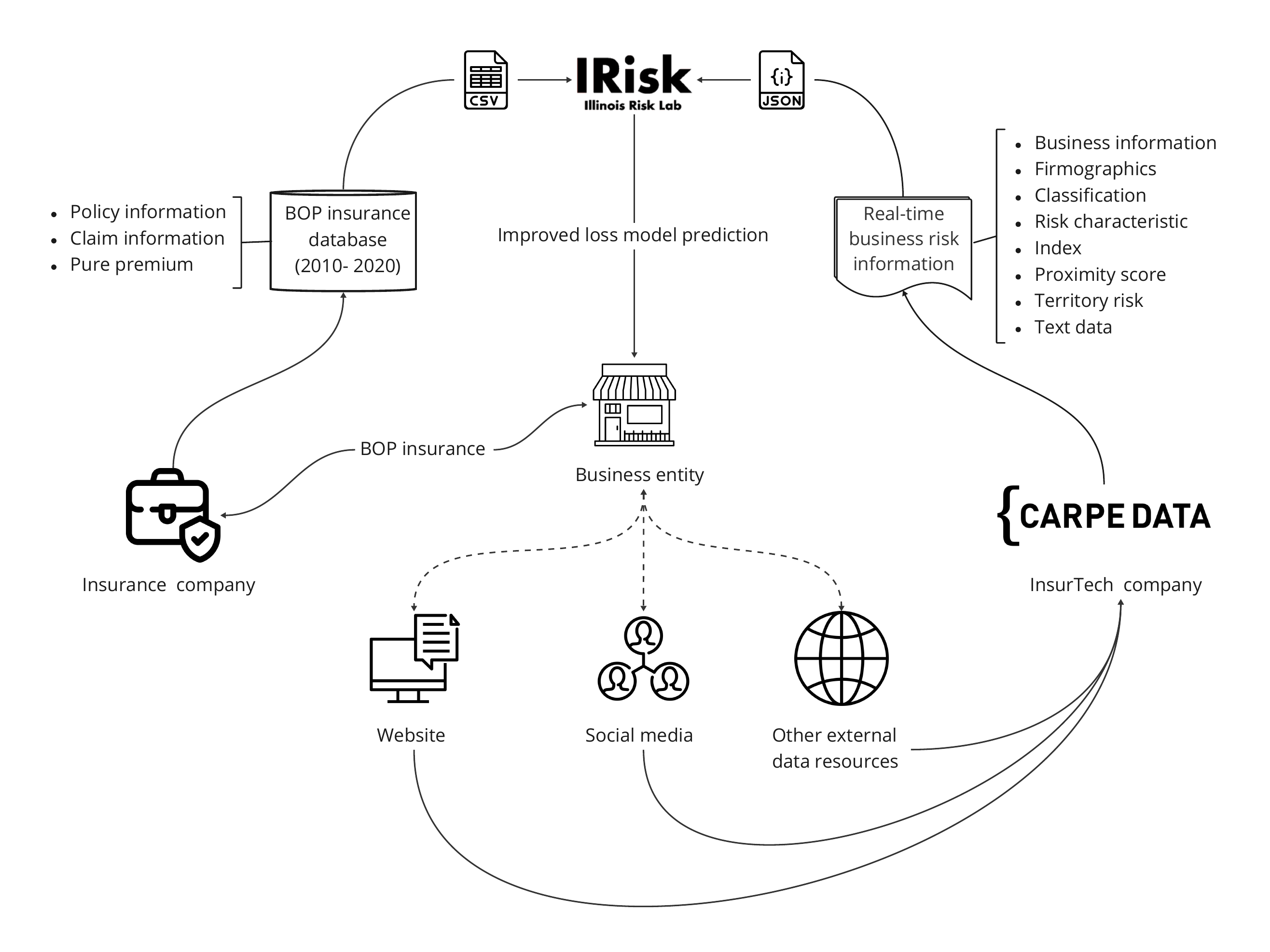}
  \caption{The flow of information from the academic-industry collaboration}
  \label{fig:structure}
\end{figure}

To address these challenges, we must pinpoint industry partners with extensive experience in business insurance loss models to conduct a comparative analysis. Additionally, collaborating with an InsurTech data vendor that offers solutions specifically designed for business insurance is indispensable. IRisk Lab\footnote{See IRisk Lab currently serves as an academic-industry collaboration hub, facilitates the integration of discovery-based learning experiences for students, and showcases state-of-the-art research in all areas of Risk Analysis and Advanced Analytics. Retrieved from \url{https://asrm.illinois.edu/illinois-risk-lab}} at the University of Illinois partnered with Carpe Data and local insurance companies to conduct joint research projects. Industry partners have provided proprietary insurance portfolio information that includes historical loss experience together with insurer in-house rating factors for exposure information, and enriched with InsurTech data. In this paper, we present a three-party research collaboration between industry and the university: the InsurTech company, the insurance companies, and the university. This collaboration is unprecedented in actuarial science. We aim to unleash the full potential of InsurTech innovation in improving business insurance loss modeling. Figure \ref{fig:structure} demonstrates the flow of information and data in this regard with IRisk Lab acting as an aggregator to be able to perform data analytics.

The rest of this paper is organized as follows. Section \ref{sec:bop} provides a brief description of business insurance, exploratory data analysis, and a mathematical justification for the advantages of bringing InsurTech innovation to the loss model.  Section \ref{sec:model} describes the calibration of the model and the results. Section \ref{sec:interpretation} provides model interpretation and suggestions for potential rating factors. We conclude in Section \ref{sec:conclude}.

\section{Opportunities for InsurTech in business insurance}\label{sec:bop}
Business insurance offers coverage designed to protect businesses and organizations of various sizes and industries from diverse risks and financial losses. It encompasses property insurance, liability insurance, business interruption insurance, commercial auto insurance, workers' compensation, and more. Additionally, bundled insurance options such as Business Owner’s Policy (BOP) and Commercial Multi-Peril (CMP) are available within the spectrum of business insurance.
\begin{table}[!ht]
\vspace*{5mm}
\centering 
\begin{tabular}{l l l r r r} 
\toprule
\multirow{2}*{Company} & \multirow{2}*{Business Line} & \multirow{2}*{Information} & \multicolumn{3}{c}{Year} \\ \cline{4-6}
& & & 2022 & 2021 & 2020 \\
\midrule
\multirow{10}*{Travelers \tablefootnote{Data extracted from: \url{https://investor.travelers.com/financial-information/annual-reports/default.aspx}}} & \multirow{2}*{CMP} &  Written Premium (in millions)  & 4,304 & 3,768 & 3,608 \\
& & Reserve for Claims and LAE (in millions) & 5,383 & 4,977 & 4,665 \\
\cline{2-6}
& \multirow{4}*{Business} & Earned Premium (in millions) & 17,095 & 15,734 & 15,294 \\
& & \( \Delta \) Loss Ratio (\%)  & -13.6  & -7.5  & -0.7  \\
& & \( \Delta \) Expense Ratio (\%) & 3.8  & 4.2  & 3.4  \\
& & \( \Delta \) Combined Ratio (\%) & -10.2  & -4.0  & 1.5  \\
\cline{2-6}
& \multirow{4}*{Personal} & Earned Premium (in millions) & 13,250 & 11,983 & 10,927 \\
& & \( \Delta \) Loss Ratio (\%) & 3.4  & -2.2  & -7.3  \\
& & \( \Delta \) Expense Ratio (\%) & -0.8  & -0.3  & -0.6  \\
& & \( \Delta \) Combined Ratio (\%) & 2.2  & -3.2  & -9.1  \\
\midrule
\multirow{10}*{Hanover \tablefootnote{Data extracted from: \url{https://investors.hanover.com/financials/annual-reports/default.aspx}}} & \multirow{2}*{CMP} & Written Premium (in millions) & 1,011 & 979 & 922 \\
& & Reserve for Claims and LAE (in millions) & 1,557 & 1,338 & 1,184 \\
\cline{2-6}
& \multirow{4}*{Business} & Earned Premium (in millions) & 1,951 & 1,811 & 1,704\\
& & \( \Delta \) Loss Ratio (\%) & -7.9  & -5.1  & -6.9  \\
& & \( \Delta \) Expense Ratio (\%) & 6.8  & 6.4  & 5.7  \\
& & \( \Delta \) Combined Ratio (\%) & -1.5  & 0.6  & -2.4  \\
\cline{2-6}
& \multirow{4}*{Personal} & Earned Premium (in millions) & 2,113 & 1,929 & 1,844 \\
& & \( \Delta \) Loss Ratio (\%) & 1.4  & -4.0  & -5.4  \\
& & \( \Delta \) Expense Ratio (\%) & 0.6  & 1.2  & 0.2  \\
& & \( \Delta \) Combined Ratio (\%) &  1.6  & -3.5  & -6.4  \\
\midrule
\multirow{10}*{Donegal \tablefootnote{Data extracted from: \url{https://investors.donegalgroup.com/financial-reporting/annual-reports-proxies}}} & \multirow{2}*{CMP} & Written Premium (in millions) & 200 & 188 & 148 \\\
& & Combined Ratio (\%) & 116.9   & 114.1  & 98.4  \\ 
\cline{2-6}
& \multirow{4}*{Business} & Earned Premium (in millions) & 510 & 468 & 413\\
& & \( \Delta \) Loss Ratio (\%) & -9.3  & -3.9  & -6.2  \\
& & \( \Delta \) Expense Ratio (\%) & 9.6  & 9.8  & 6.0  \\
& & \( \Delta \) Combined Ratio (\%) & 1.0  & 5.2  & -1.0  \\
\cline{2-6}
& \multirow{4}*{Personal} & Earned Premium (in millions) & 312 & 307 & 329 \\
& & \( \Delta \) Loss Ratio (\%) & -5.4  & -7.7  & -10.6  \\
& & \( \Delta \) Expense Ratio (\%) & 5.9  & 2.1  & 4.8  \\
& & \( \Delta \) Combined Ratio (\%) &  0.1 & -5.3 & -3.4 \\
\midrule
\midrule
\multirow{3}*{Baseline \tablefootnote{Adapted from Federal Insurance Office, U.S. Department of the Treasury, Annual Report on the Insurance Industry (2023), 44, \texttt{https://home.treasury.gov/system/files/311/FIO\%20Annual\%20Report\%202023\%209292023.pdf}.
}} & \multirow{3}*{P\&C} & Loss Ratio (\%) & 76.4   & 72.5  & 70.1  \\
& & Expense Ratio (\%) & 25.9   & 26.5  & 27.5  \\
& & Combined Ratio (\%) & 102.7  & 99.7  & 98.8\\
\bottomrule
\end{tabular}
\caption{Earned Premium and Claim Payout Information by Insurers}
\label{tab:insurer}
\end{table}

BOP aims to protect small- or medium-sized business owners from potential business risks through a bundle of insurance policies with various coverages. See \cite{Naylor2017}. Most BOP bundles contain essential coverage of property insurance and liability insurance, while additional coverage, such as commercial auto insurance, commercial crime insurance, business interruption insurance, and other business insurance policies, can be included through customization or negotiation. CMP is often more suitable for larger businesses with complex insurance requirements, given its flexibility and ability to address a wide range of risks. The underwriting procedure for business insurance necessitates a more thorough analysis and assessment, leading to increased underwriting expenses. Delivering a heightened level of customization and personalized service could incur higher administrative costs. Additionally, the distribution channels for business insurance often include more intermediaries, such as agents and brokers, which affects the expense ratio. We select three representative insurance companies, namely, Travelers for large size, Hanover for midsize, and Donegal for small size, offering a diverse view of insurers based on their scale. Their respective market shares and detailed loss summaries, extracted from their annual reports, are presented in Table \ref{tab:insurer}. Here, $\Delta$ represents the difference between the operational ratios (Loss Ratio, Expense Ratio, Combined Ratio) of the selected companies and the average ratios in the P\&C sector. It is evident from the table that business insurance exhibits a higher expense ratio, approximately 4-5\%, in contrast to personal insurance. The business insurance expense ratio is around 30\%, nearly half of the corresponding loss ratio. Furthermore, smaller insurers display a higher expense ratio, especially in the context of business insurance, where the difference is more pronounced compared with larger insurers. Despite human intervention in the underwriting process, insurers fall short of gathering sufficient risk factors essential for accurate pricing, resulting in elevated loss ratios. This trend is commonly observed in bundled business insurance across the industry. Regardless of the size of the insurers, all selected companies show that for bundled business insurance such as CMP, the reserve for claims and loss adjustment expense (LAE) is higher than the written premium. In the case of our insurer partner, we observe the Actual/Expected (A/E) ratio of 107.52\%, representing the ratio of actual losses to those predicted by the insurer's in-house loss modeling, for the entire BOP bundle.

The current circumstances clearly require enhancement, and there is a demand for effective solutions. Fortunately, InsurTech can deliver a customized underwriting system with dynamic programmable rules. The InsurTech-enhanced features, as another typical InsurTech application, empower underwriters and claims adjusters with access to enhanced information about policyholders and claims, facilitating more informed decision-making. For instance, a customized underwriting system can automatically pre-fill information for the standard underwriting process, saving time for underwriters, which would otherwise be spent on communicating with policyholders or manually searching for information from various sources. This efficiency has the potential to drive costs down and even enable process automation for smaller policies such as BOP without the need for direct intervention from underwriters. By harnessing new and multiple external data sources, including social media data and online content, analytics are empowered and accurate information is gained on risks and policyholder behavior. Furthermore, data enrichment and improved internal data management contribute to more precise pricing and the formulation of data-driven underwriting strategies for traditional insurance companies.

In the subsequent section, we illustrate our datasets gathered from industry partners; an anonymous insurance company shares its proprietary loss experience over the past decade, which is an entire database created to make current rate plans for BOP, along with partial rating factors for in-house loss modeling. The InsurTech company, Carpe Data, provides dynamic and real-time information from emerging public data sources that illuminate every aspect of a business, including operations, products, services, and reviews. It offers potential risk factors by connecting the different risk profiles. In summary, the response variable is the BOP loss cost during the observation period. For predictive features, we have two sources: the insurance company provides policy information used as rating factors, and Carpe Data offers policyholders' risk factors.

\subsection{Insurer's BOP policy information}\label{subsec::bop_data}
In our BOP dataset, there are three types of coverage: Business Building (BG), Business Personal Property (BP), and Liability (LIAB). BG explicitly covers the loss related to the buildings in which the business operates, owned or rented by the business owners. BP covers the risks of potential loss, damage, and liability issues for the property of policyholders that has a business use. The losses covered by BG and BP are discovered and covered quickly (short-tail) with relatively minor claims, which are often capped by the value of the properties. LIAB covers the risks of potential legal liability for property damage, bodily injury, and other losses caused by policyholders to a third party, for example, medical payments to customers who are injured while on business property, defense costs, damages for failure to or improperly rendering professional service, and accidental pollution. Compared to BG and BP, the losses covered by LIAB are usually long-tail risks, where the report and settlement can take a long time, and the frequency and severity of claims vary depending on industry and type of incident.
\begin{table}[!ht]
\vspace*{5mm}
\centering 
\begin{tabular}{ L{3.5cm} L{4.5cm} L{5.7cm}} 
\toprule
Category & Features & Details \\
\midrule
\multirow{3}*{Policy Information} 
& Policy Year  & From 2010 to 2020 \\ \cline{2-3}
& Earned Exposure &  \\ \cline{2-3}
& Coverage Limit &  \\ \cline{2-3}
& Coverage Type &  Building (BG), Business Personal Property (BP), Liability (LIAB)  \\ \cline{2-3}
& Exposure Base & LOI, Annual Gross Sales, Annual Payroll\\ \cline{2-3}
& Risk Type & Apartment, Condo/Office, Contractors, Convenience, Distributor, Fast Food, Motel, Office, Other, Restaurant, Retail, Self-Service \\ 
\midrule
\multirow{2}*{Loss experience}
& Observed Loss Cost  & \\ \cline{2-3}
&Insurance Company's In-house Model Loss Cost &  \\ \bottomrule
\end{tabular}
\caption{Policy information and loss experience provided by the insurance company}
\label{tab:insurance-features}
\end{table}

As earlier mentioned, the response variable is the BOP loss cost observed from 2010 to 2020. Table \ref{tab:insurance-features} outlines the response variable together with the features provided by the insurance company. It is worth mentioning that the insurance company also attaches a pure premium for each policyholder, which is the insurance company's in-house model loss cost. Due to privacy concerns, we do not have all the rating factors that are used in the in-house model. However, this does not preclude us from directly comparing our results with the in-house model results, which will be discussed in Table \ref{tab:VM-result} of Section \ref{sec:interpretation}. 

\begin{figure}[!ht]
  \centering
  \includegraphics[width= 0.6\linewidth]{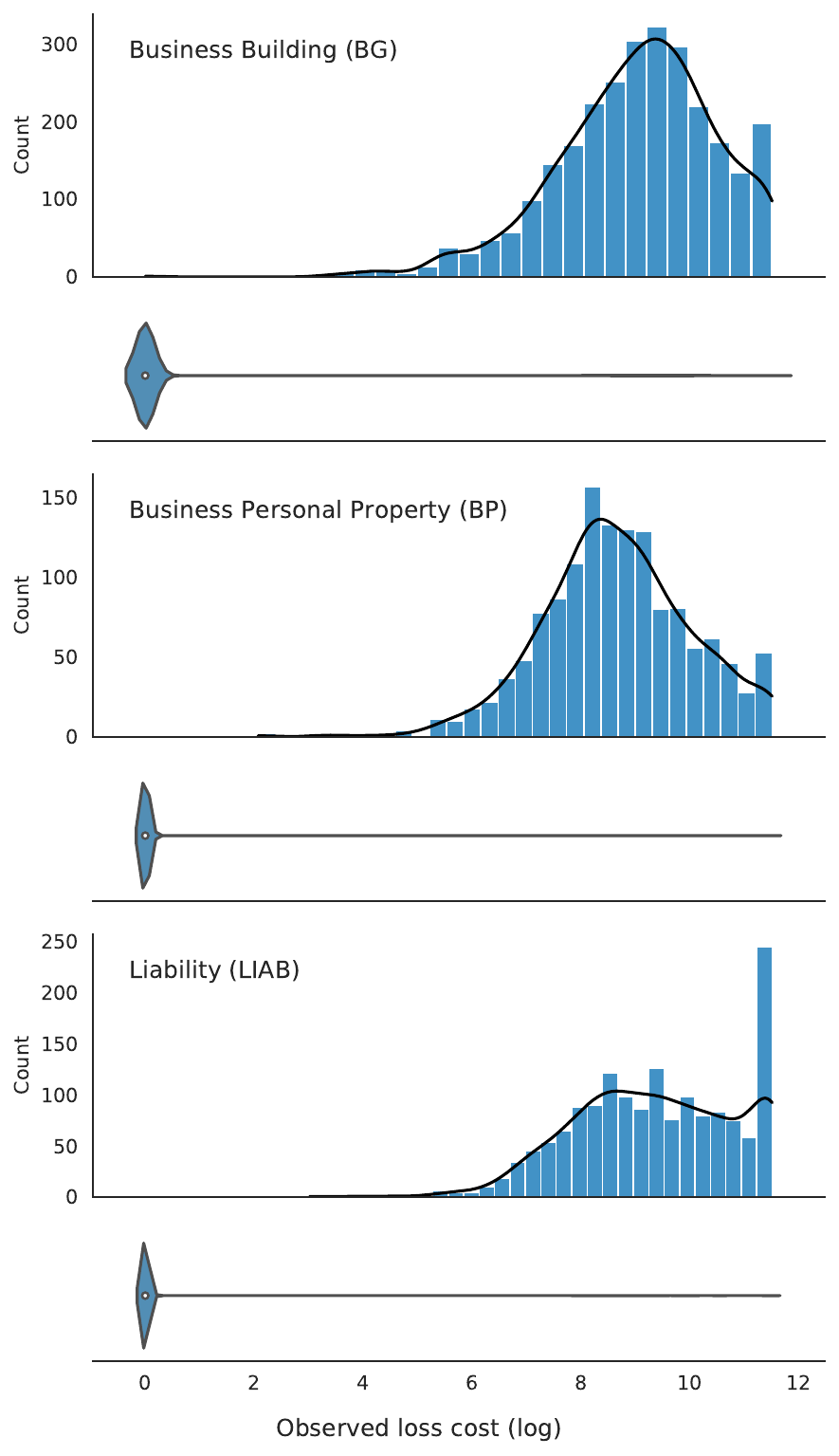}
  \caption{The distribution of the observed loss cost by coverage type}
  \label{fig:obsloss}
\end{figure}

Figure \ref{fig:obsloss} shows the distribution of our response variable or the observed loss cost, by the three coverage groups. The violin plots show the kernel density for the logarithm of the observed loss cost plus one. We can observe that the majority of policies have zero loss costs, meaning that most policyholders do not have claims submitted. Based on the shapes of the distributions, we see that the loss cost of the LIAB group is centered around zero with the most significant proportion. Specifically, 96.50\% of the policies of the BG group, 98.99\% of the policies of the BP group, and 99.20\% of the policies of the LIAB group have zero claims. In addition, we present histograms showing how the logarithm of the positive observed loss cost plus one is distributed. We can deduce that the median value of the observed loss from the LIAB group (around $e^{12}$) is higher than those of the other two groups. The other two groups, BG and BP, are around $e^9$ and $e^{10}$, respectively. Finally, we observe that some policyholders tend to report claims more often or to report more expensive claims than others. In practice, most insurance portfolios are heterogeneous because they are usually a mix of policyholders with different risk levels. This is especially true for BOP with three different coverages. After basic exploratory data analysis and comprehensive model experiments, we decide to model these three coverage groups separately.

Next, we focus on exploratory data analysis and get a first impression of how features influence the response variable. Clearly, due to space constraints, we can show only some of the representative features. However, we focus on some of the features that appear to be worth discussing after model fitting. 

\begin{figure}[!ht]
\centering
\begin{subfigure}{.48\textwidth}
\centering
\includegraphics[width=0.9\linewidth]{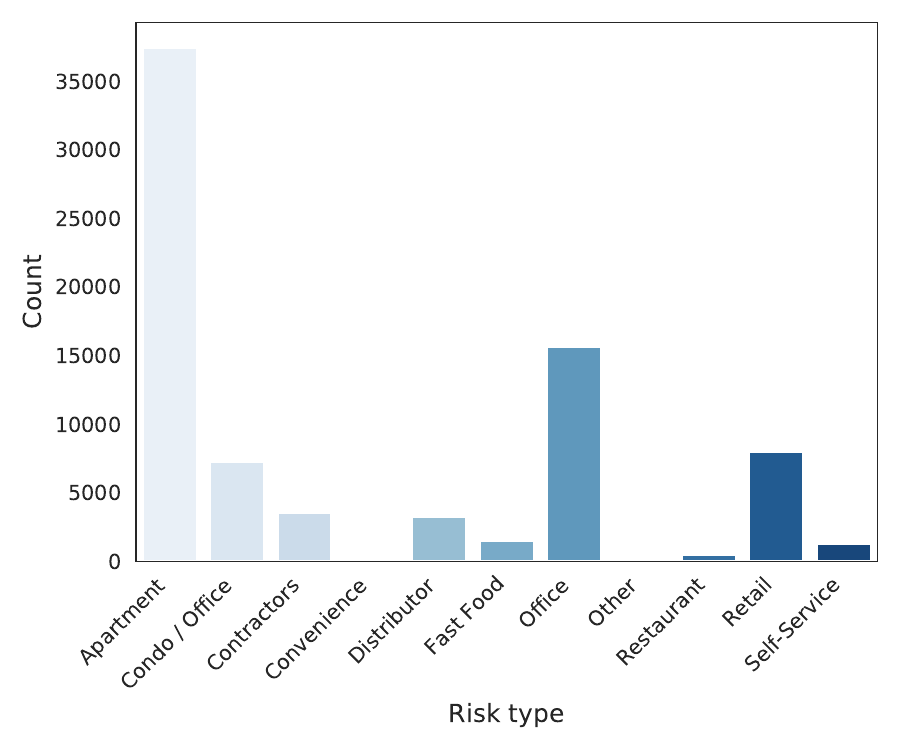}
\caption{Histogram of risk type}
\label{fig:hist_risktype}
\end{subfigure}
\begin{subfigure}{.48\textwidth}
\centering
\includegraphics[width=0.9\linewidth]{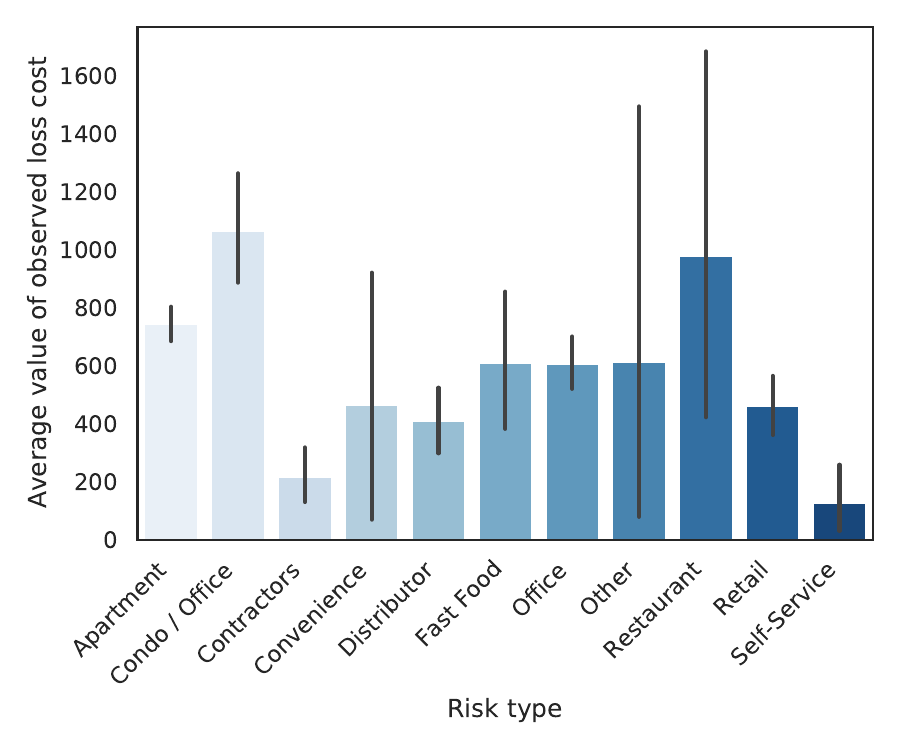}
\caption{Barplot of average observed loss cost by risk type}
\label{fig:bar_risktype_loss}
\end{subfigure}
\begin{subfigure}{.48\textwidth}
\centering
\includegraphics[width=0.9\linewidth]{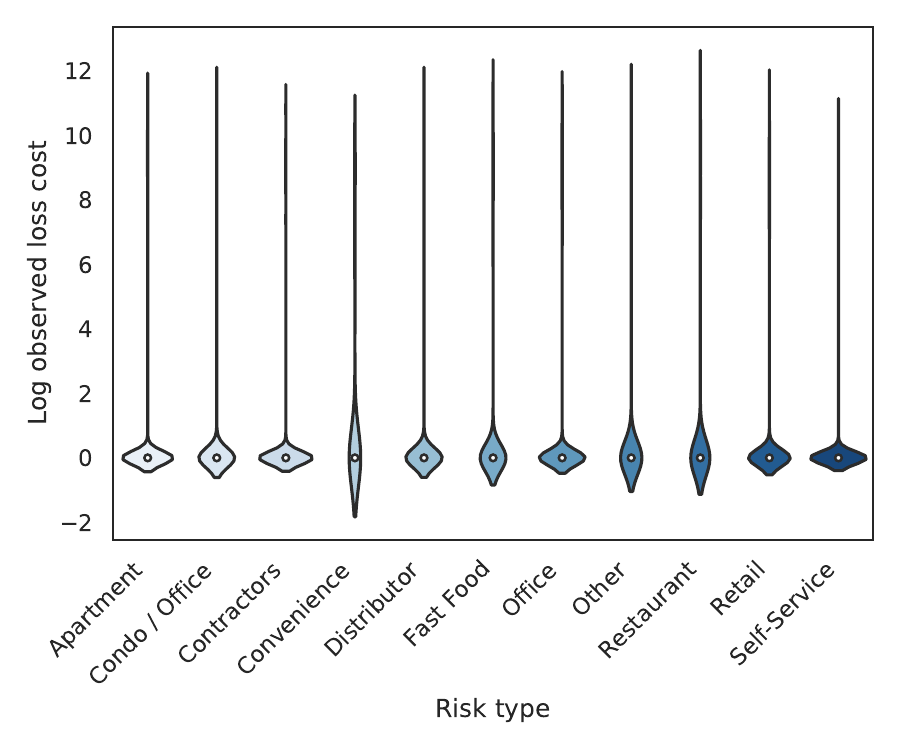}
\caption{Violin plot of log value of observed loss cost by risk type}
\label{fig:violinplot_risktype_loss}
\end{subfigure}
\begin{subfigure}{.48\textwidth}
\centering
\includegraphics[width=0.9\linewidth]{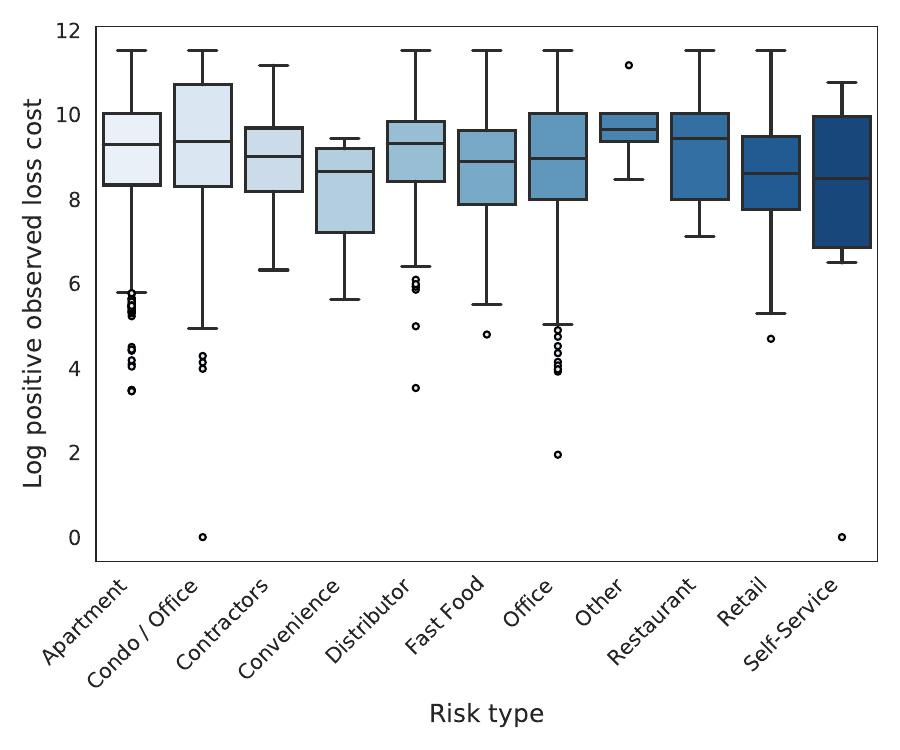}
\caption{Box plot of log value of positive observed loss cost by risk type}
\label{fig:boxplot_risktype_loss_nonzero}
\end{subfigure}
\caption{Risk type}
\label{fig:risktype}
\end{figure}

Let us pick one rating factor, the risk type, from the policy information provided by the insurance company to show one-way analyses. We note that in this and subsequent preliminary analysis, the observed loss cost and features analyzed are for BG coverage only, due to the limitation of paper length. Figure \ref{fig:risktype} demonstrates eleven risk types in total, ranging from \textit{Apartment} to \textit{Self-Service}. Subfigure \ref{fig:hist_risktype} shows the number of policies based on the different types of risk. We can see that the \textit{Apartment} risk type has the most significant amount, followed by the \textit{Office} risk type and the \textit{Retail} risk type, whereas the \textit{Convenience} risk type has the least amount. Subfigure \ref{fig:bar_risktype_loss} depicts the average value of the observed loss cost for each different type of risk. The black line shows the standard deviation of the observed loss cost. On average, the \textit{Condo/Office} risk type BOP policies have the highest observed loss cost, whereas those of the \textit{Self-Service} risk type have the lowest. Subfigure \ref{fig:violinplot_risktype_loss} presents the violin plots of the logarithm of the observed loss cost plus one by the different types of risk. The shapes of the distributions reveal that the loss cost from the \textit{Convenience} risk type is the most dispersed, whereas the \textit{Self-Service} risk type is the most concentrated. To highlight the differences in the observed loss cost by different types of risk, we also show boxplots of the logarithm of the positive observed loss cost in Subfigure \ref{fig:boxplot_risktype_loss_nonzero}. It shows that the median values across all types of risk are relatively similar, floating around $8000$ ($e^9$); however, the dispersion for the different types of risk varies significantly. This is yet another evidence of the complexity of BOP across multiple industries and under extensive risk exposure.

\subsection{Data from InsurTech}
\begin{table}[!ht]
\vspace*{5mm}
\centering
\begin{tabular}{ L{3.8cm} L{5.5cm} L{.38\textwidth}}
\toprule
Category & Explanation & Examples \\
\midrule
Business Information & Basic information concerning the operation of businesses including address, year founded, operating hours, etc. & \textit{coordinates, state, street\_type, type\_of\_address, is\_franchise, is\_home\_business, operating\_year, language\_spoken, opening hours} \\ \hline
Firmographics & Characteristics to segment prospect business including the type of business, the number of employees, the sales ranges, etc. & \textit{business\_size, credit\_score, company\_type, sales\_range, revenue\_range} \\ \hline
Classification & Categorization of a business & \textit{category, segment, segment\_details, NAICS code} \\ \hline
Risk Characteristics & Indicator features that identify current attributes relating to the potential risks of a business 
& \textit{Alcohol, Chemical Application, Fireplace, Gasoline, Heavy Construction, Outdoor Heaters, Raw Seafood Served, Tobacco, Work at Heights} \\ \hline
Index & A suite of indexes on a 1 – 5 scale targeting dimensions of risk that can be tuned by segment and location & \textit{Customer Rating, Visibility, Reputation, Health \& Sanitation, Maintenance \& Condition.} \\ \hline
Proximity Score & Proximity scores identify risks associated with the surrounding businesses that may impose on policyholders in the immediate vicinity of that risk  & \textit{Negative Keywords, Combustibles, Entertainment, Traffic} \\ \hline
Territory Risk &  Density scores of risks within a zip code area, engineered from collected information based on the location of a business & \textit{TERRITORY.a1 to TERRITORY.m1}
\textit{TERRITORY.a2 to TERRITORY.m2}
\\ \hline
Text Data & (1) Webpage content crawled from a business' website; and (2) Customer reviews of a business from multiple data sources, including but not limited to the content of the review, star rating, number of likes, the date of posting, and the source of the reviews. & \textit{Webpage content: content, title, url} \textit{Customer reviews: content, likes, dislikes, stars, saves, language} \\  
\bottomrule
\end{tabular}
\caption{Features from InsurTech company}
\label{tab:InsurTech-features}
\end{table}
The potential of InsurTech innovation to help the insurance company model these extensive risks has never been greater. However, unlike most academic toy datasets, we cannot safely assume that our real-life dataset is reliable and accurate. First, we need to carefully check the entity resolution, which evaluates if the InsurTech company is able to match the correct policyholder based on the limited information, including address, business name, and others. Second, we need to perform data engineering work, including extracting structured datasets from the raw data files provided in nested JSON (JavaScript Object Notation) format with hierarchical structures; cleaning dirty data, which includes missing data, inconsistent format, and human input errors; creating useful features from various data types, including text data. Third, we need to set up our relational database that connects the InsurTech features. The final version of the combined dataset has $825,622$ observations and $596$ features. Table \ref{tab:InsurTech-features} broadly summarizes various categories of InsurTech features with explanations and examples. 

The InsurTech features of the classification category have information similar to the risk types mentioned in Section \ref{subsec::bop_data}, but provide a more granular categorization of a business. There are $24$ segments that provide categorization of a business, varying from \textit{Accountants \& Financial Services} to \textit{Wholesale \& Distributors}. Additionally, companies can operate in multiple lines of business, thus falling into multiple segments. Carpe Data utilizes various classifiers to assign a segment to a business, and segment details provide the proportion of total votes of a certain segment for a business. For example, one of the policyholders is a grocery store run by a family farm that sells and delivers meat to the surrounding community. It belongs to the \textit{Retail} risk type based on insurance policy information and is also segmented as \textit{Retail} (with a majority vote of $50\%$) from the InsurTech aspect. Furthermore, based on segment details information, it is also categorized as \textit{Wholesale \& Distributors} ($20\%$, sells and delivers meat to the surrounding company), \textit{Agriculture}, \textit{Forestry}, \textit{Fishing \& Hunting} ($10\%$, farm), \textit{Food Services \& Drinking Places} ($10\%$, grocery has food service), and \textit{Manufacturing} ($10\%$, produce dairy products). Despite the overlap between risk types from in-house rating factors and segment information from InsurTech innovations, InsurTech features can cross-validate and enhance internal rating factors.

\begin{figure}[ht!]
\centering
\begin{subfigure}[t]{.5\textwidth}
\centering
\includegraphics[width= 0.9\linewidth]{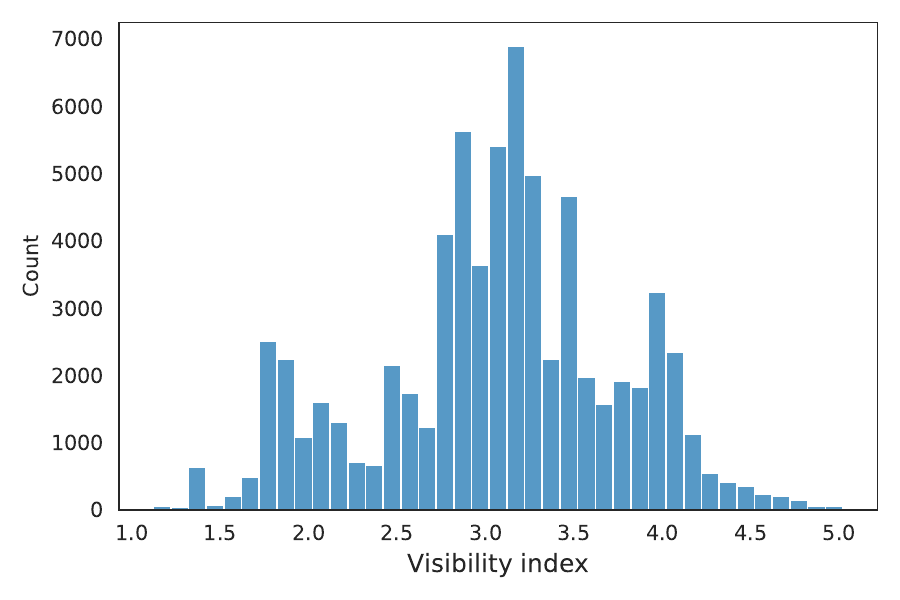}
\caption{Histogram of visibility index}
\label{fig:hist_visibility}
\end{subfigure}%
\begin{subfigure}[t]{.5\textwidth}
\centering
\includegraphics[width=0.9\linewidth]{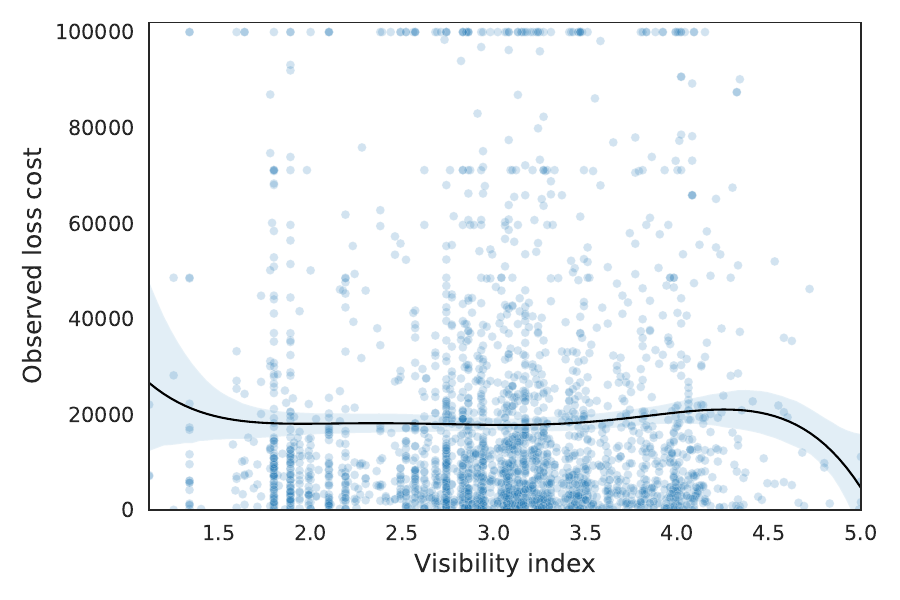}
\caption{Scatter plot of visibility index and observed loss cost with a nonlinear regression line}
\label{fig:scatter_visibility_mean_loss_nonlinear}
\end{subfigure}
\caption{Visibility index}
\label{fig:visibility}
\end{figure}

We now focus on exploratory data analysis for some of the representative InsurTech features mentioned in Table \ref{tab:InsurTech-features}. Indexes are a suite of tools that measure the risks faced by a business. Among them, the visibility index measures a business entity's visits and online presence based on social media, visitor reviews, and marketing information. Figure \ref{fig:visibility} illustrates the visibility index and its univariate relationship with the observed loss cost. Subfigure \ref{fig:hist_visibility} shows the distribution of the visibility index, which ranges from 1.0 to 5.0 and peaks at about 3.2. This score helps measure the absolute risk of a business, the risk relative to other businesses, and the change in risk over time. Subfigure \ref{fig:scatter_visibility_mean_loss_nonlinear} presents a scatter plot of the visibility index and the positive observed loss cost with a nonlinear regression line. The trend of the regression line shows that the loss costs are higher when the visibility index is lower and vice versa. Visibility is an important tool for identifying potential customers to improve business performance. 

\begin{figure}[ht!]
\centering
\begin{subfigure}[t]{.5\textwidth}
\centering
\includegraphics[width= 0.9\linewidth]{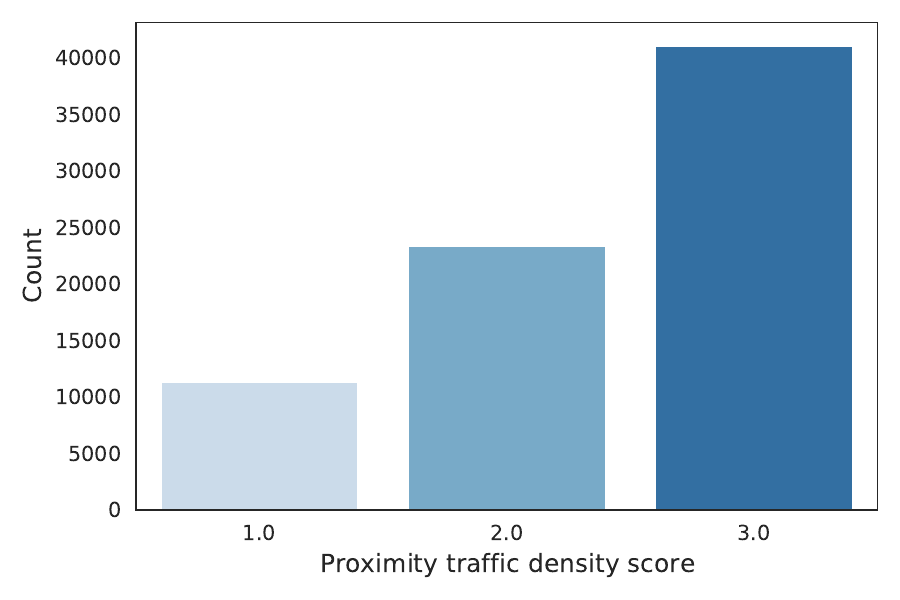}
\caption{Histogram of traffic density score}
\label{fig:hist_traffic_density}
\end{subfigure}%
\begin{subfigure}[t]{.5\textwidth}
\centering
\includegraphics[width= 0.9\linewidth]{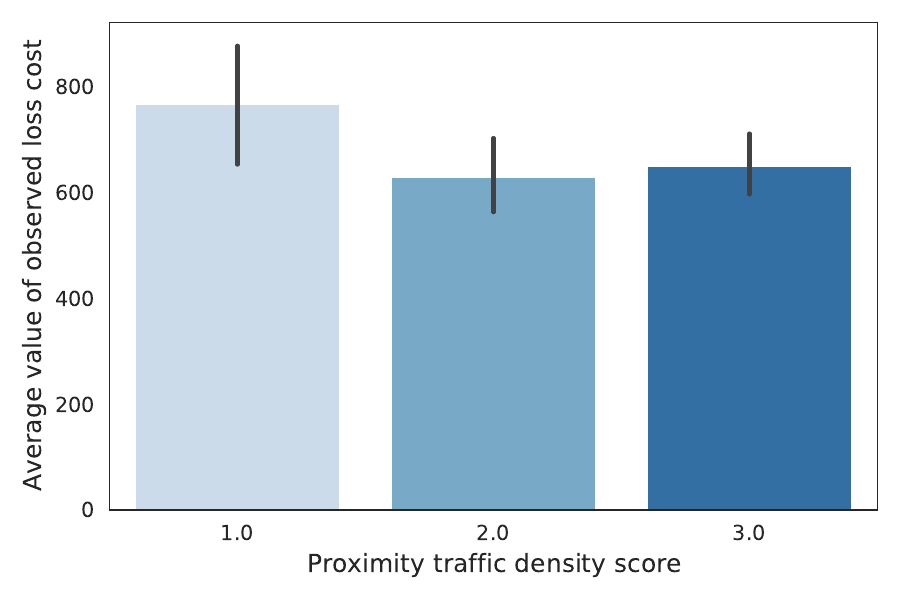}
\caption{Barplot of average value of observed loss cost by traffic density score}
\label{fig:bar_traffic_density_loss}
\end{subfigure}
\begin{subfigure}[t]{.5\textwidth}
\centering
\includegraphics[width=0.9\linewidth]{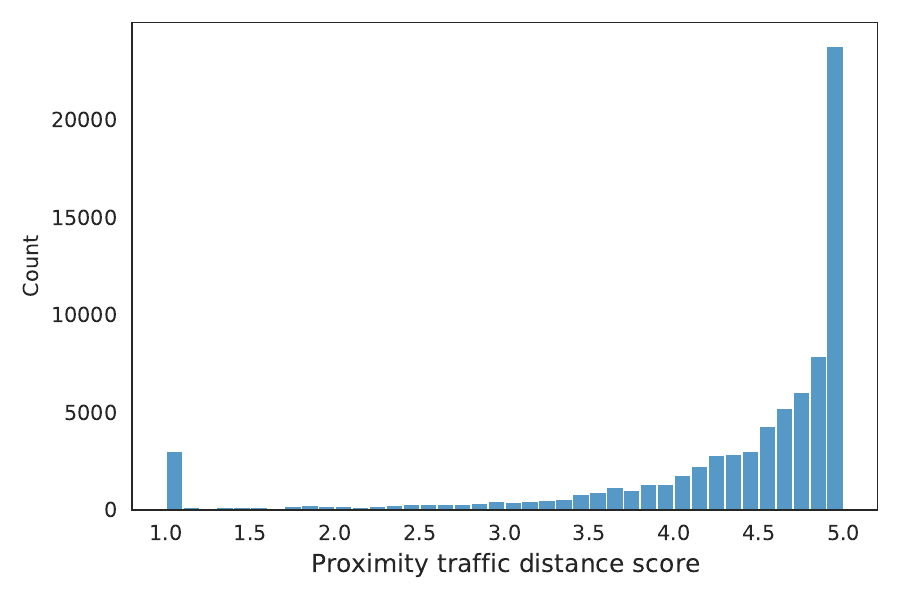}
\caption{Histogram of traffic distance score}
\label{fig:hist_traffic_distance}
\end{subfigure}%
\begin{subfigure}[t]{.5\textwidth}
\centering
\includegraphics[width=0.9\linewidth]{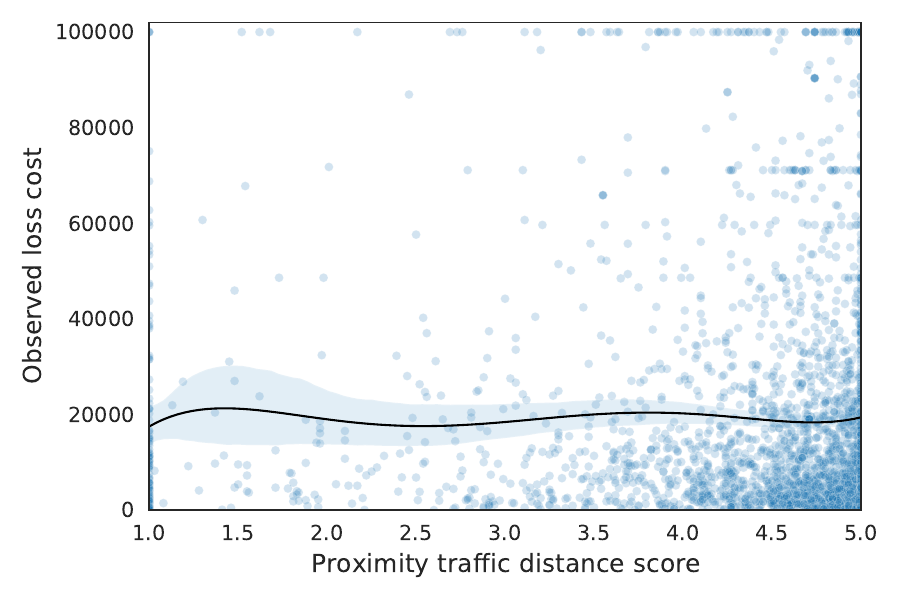}
\caption{Scatter plot of traffic distance score and observed loss cost with a nonlinear regression line}
\label{fig:scatter_traffic_distance_loss}
\end{subfigure}
\caption{Proximity Traffic Scores}
\label{fig:traffic}
\end{figure}

Proximity scores identify risks associated with the surrounding businesses that can expose policyholders to additional risks in their immediate vicinity. Three groups of proximity scores are considered in our datasets: combustibles, entertainment, and traffic. Each group has two scores: the density score and the distance score. The density score is calculated over a certain range by counting the total number of risks and normalized to a value that is an integer from 1 to 3, where 1 indicates high risk and 3 indicates low risk; the distance score measures the nearby risks as a function of distance. It is a float number between 1.0 and 5.0, with a higher number indicating a lower degree of risk. Figure \ref{fig:traffic} illustrates one group of proximity scores, which is proximity traffic, and shows how it relates to the observed loss cost. As we mentioned, there are two types of scores for proximity traffic, the density score and the distance score. Subfigure \ref{fig:hist_traffic_density} compares the number of policies with different density scores, showing that the number of policies increases with the density score, with a density score of 3 having the highest number of policies. Similarly, Subfigure \ref{fig:hist_traffic_distance} illustrates a more granular view of how distance scores are distributed, and we observe a consistent trend that there are more observations with a distance score close to 5. Subfigure \ref{fig:bar_traffic_density_loss} shows that there has been a steady decline in the average value of the observed loss cost as the density score of proximity traffic increases, which matches the intuition that a lower surrounding risk implies a lower claim loss. Additionally, businesses with a density score of 1 have the highest variability in observed loss cost. Subfigure \ref{fig:scatter_traffic_distance_loss} presents a scatter plot of the distance score and the observed loss cost with a nonlinear regression line to get a general sense of the relationship. We see that this nonlinear regression line has a nearly constant trend, but the confidence interval becomes narrower with more observations, especially as the proximity traffic distance score reaches 5. Usually considered as a feature that may indicate accessibility, proximity to traffic can be a complex variable with an unclear impact on business performance, as its effect can vary widely depending on the type of business.

\begin{figure}[ht!]
\centering
\begin{subfigure}[t]{.5\textwidth}
\centering
\includegraphics[width= 0.9\linewidth]{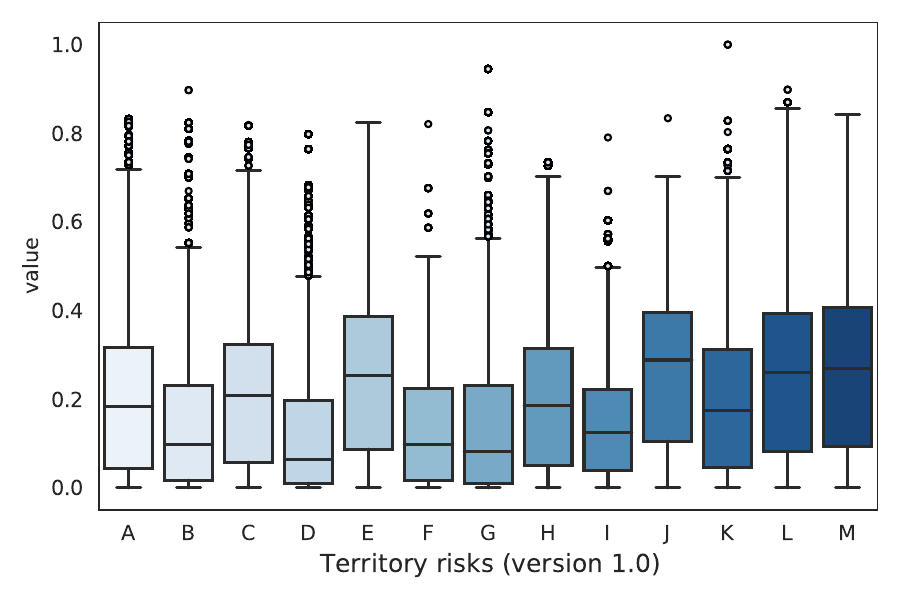}
\caption{Boxplot of territory risk (version 1.0)}
\label{fig:boxplot_group_2020}
\end{subfigure}%
\begin{subfigure}[t]{.5\textwidth}
\centering
\includegraphics[width= 0.9\linewidth]{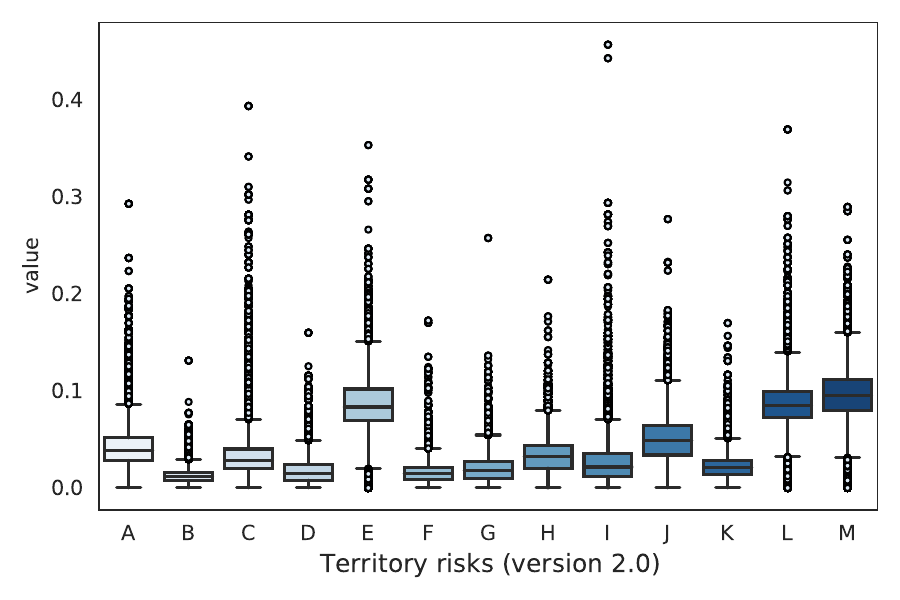}
\caption{Boxplot of territory risk  (version 2.0)}
\label{fig:boxplot_group_2021}
\end{subfigure}
\begin{subfigure}[t]{.5\textwidth}
\centering
\includegraphics[width=0.9\linewidth]{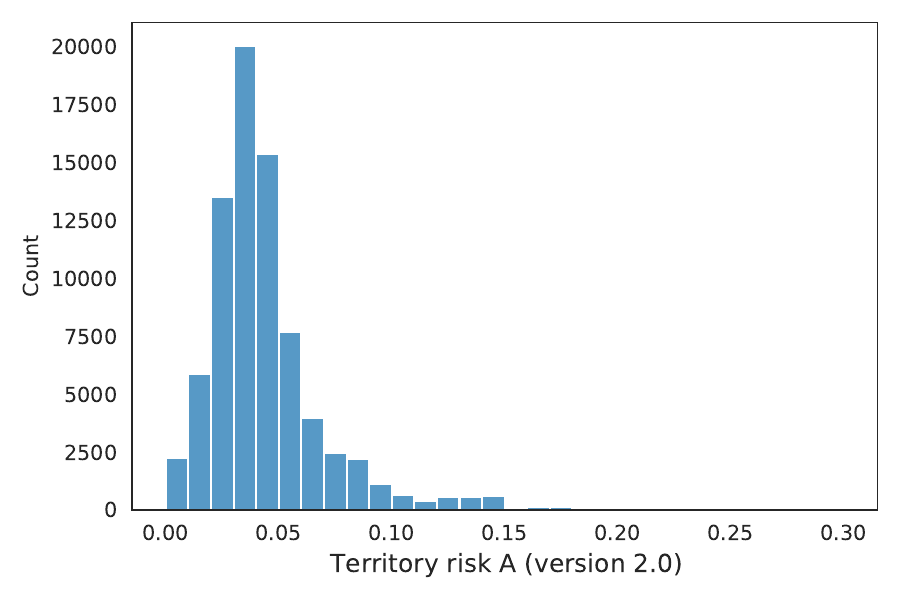}
\caption{Histogram of territory risk A (version 2.0)}
\label{fig:hist_group_1_2021}
\end{subfigure}%
\begin{subfigure}[t]{.5\textwidth}
\centering
\includegraphics[width=0.9\linewidth]{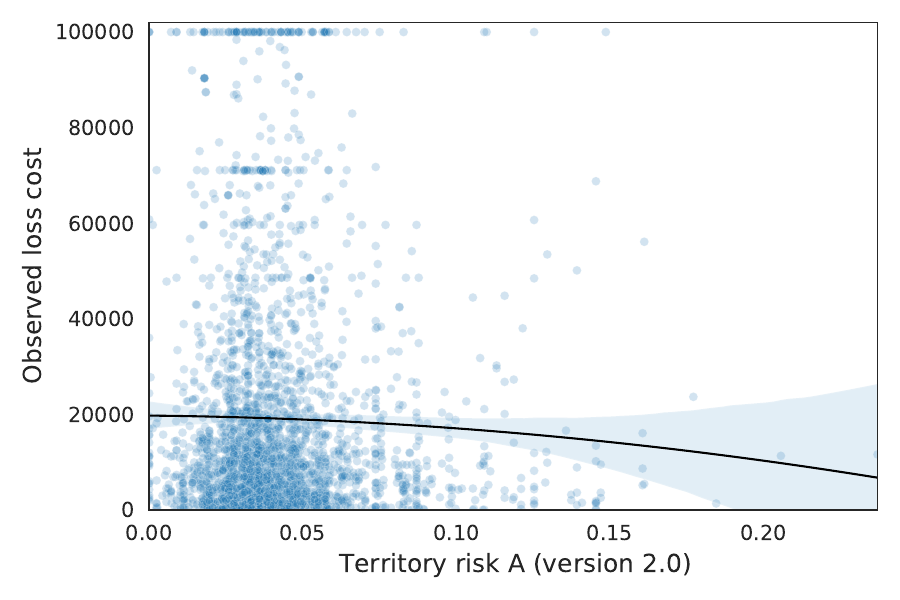}
\caption{Scatter plot of territory risk A (version 2.0) and observed loss cost with a nonlinear regression line}
\label{fig:scatter_group_1_2021_loss}
\end{subfigure}
\caption{Territory risk}
\label{fig:group_features}
\end{figure}

Territory risk measures the cluster of risks within the zip code level area, and it is engineered from the business risk characteristics defined in Table \ref{tab:InsurTech-features}. Since it is determined based on the zip code level, it can readily be incorporated with existing location-based rating factors for insurance company use. There are 13 clusters of risks, named risk A to risk M, that focus on different aspects of the business risk characteristics. For example, one of the clusters may focus on fire-related risk characteristics concerning potential fire hazards in this territory. Based on different feature engineering techniques, two versions of territorial risk, namely version 1.0 and version 2.0, are used in our modeling. Both versions are continuous and range between 0 and 1 for version 1.0 and between 0 and 0.5 for version 2.0. For each version, a higher value indicates a lower risk. Subfigure \ref{fig:boxplot_group_2020} and Subfigure \ref{fig:boxplot_group_2021} show the boxplots of territorial risk with two different versions. We can see that the features of version 1.0 are more dispersed than those of version 2.0. Subfigure \ref{fig:hist_group_1_2021} shows the right-skewed distribution of territorial risk feature A, ranging from 0 to 3.0, with the most common value of 0.04. Subfigure \ref{fig:scatter_group_1_2021_loss} presents a scatter plot of the territorial risk feature A (version 2.0) and the observed loss cost. 

\subsection{Information gained from InsurTech}
Based on datasets obtained from our industry partners, we focus on improving BOP loss models by leveraging InsurTech innovations. P\&C insurance carriers started to apply modern predictive modeling in the early 2000s. Meanwhile, insurance companies began to adopt analytics with large insurance datasets for various applications. For example, actuarial ratemaking, or actuarial pricing, described in \cite{denuit2007actuarial}, refers to the process during which actuaries quantify the insurance products for the underlying risks covered by insurance policies. The process of ratemaking targets to accurately predict future claim losses by policyholders based on historical loss data.
One of the most significant underpinnings of the ratemaking process is data. The accuracy and fairness of the premium rates that are ultimately determined depend largely on the quality and quantity of available data. Therefore, it is natural to look for good-quality external data to enhance the underlying predictive models. 
In addition, it is worth mentioning that insurance is highly regulated and business insurance premiums must comply with the laws set by local regulators. Such regulation usually serves two purposes: first, to ensure the insurer's solvency, and second, to protect customers by restricting unfair rating factors. 
Hence, in this paper, our goal is two-fold: mining predictive risk characteristics from InsurTech data that improve insurance in-house loss model, and explaining these risk characteristics using interpretable machine learning techniques. We also try to further propose potential rating factors for business insurance.

Let us define $\textbf{X}^{IH}$ as in-house rating factors used by the insurance company, such as policy information, basic policyholder information, and conventional risk characteristics. $\textbf{X}^{IT}$ denotes risk factors generated by InsurTech innovations, such as enhanced policyholder information, firmographics and territory information, and related social media information. We assume $\textbf{X}^{IH}$ and $\textbf{X}^{IT}$ are subsets of $\textbf{X}^{W}$, which is a set of ground-truth risk factors that perfectly define the underlying risk. Obviously, $\textbf{X}^{W}$ is unobtainable and unrealistic in a real-life setting. We further assume that there is no overlap between $\textbf{X}^{IH}$ and $\textbf{X}^{IT}$ without loss of generality. 
Indeed, there may be some overlap in the information between insurance in-house rating factors and InsurTech risk factors. However, the choice to utilize overlapping data usually entails a careful consideration of the trade-offs between costs and benefits. Such overlapping information can serve as a cross-reference to validate insurance in-house rating factors, potentially paving the way for further improvements.

Let $Y$ be the claim amount, which is our response variable, and the variance of $Y$ can be decomposed according to the law of total variance as
$$
\operatorname{Var}(Y) = \operatorname{E}[\operatorname{Var}(Y \mid \textbf{X}^{IH})] + \operatorname{Var}(\operatorname{E}[Y \mid \textbf{X}^{IH}]),
$$
where $\operatorname{E}\left[Y \mid \textbf{X}^{IH}\right]$ is the current pure premium based on the insurance in-house loss model. The insurer believes that given current rating factors, $\operatorname{Var}(Y \mid \textbf{X}^{IH})$ is considered random as white noise, as the variability in claim amounts is solely due to chance. However, this belief can be arguable, especially when there is a large information discrepancy between $\textbf{X}^{IH}$ and $\textbf{X}^{W}$. The insurer must carefully monitor this variance in the presence of systematic drift. The second term can be deciphered from the portfolio perspective, i.e., based on the current pure premium, $\operatorname{E}\left[Y \mid \textbf{X}^{IH}\right]$, it is the variance of the current pure premium that depends on the portfolio composition. If the current pure premium is not precise enough, then adverse selection could occur, and the change in the portfolio composition could generate systematic losses for the insurer.

The first term can be further decomposed, using \cite{bowsher2012identifying}, by bringing InsurTech risk factors, $\textbf{X}^{IT}$, as follows:
$$
\operatorname{Var}(Y) = \operatorname{E}\left[\operatorname{Var}\left(Y \mid \textbf{X}^{IH}, \textbf{X}^{IT}\right)\right] + \operatorname{E}[\operatorname{Var}(\operatorname{E}\left[Y \mid \textbf{X}^{IH}, \textbf{X}^{IT}\right] \mid \textbf{X}^{IH})] + \operatorname{Var}(\operatorname{E}\left[Y \mid \textbf{X}^{IH}\right]),
$$
where $\operatorname{E}\left[Y \mid \textbf{X}^{IH}, \textbf{X}^{IT}\right]$ is the updated pure premium by leveraging InsurTech innovation. Evidently, $\operatorname{E}\left[\operatorname{Var}\left(Y \mid \textbf{X}^{IH}, \textbf{X}^{IT}\right)\right]$ is smaller than $ \operatorname{E}[\operatorname{Var}(Y \mid \textbf{X}^{IH})]$ and could be closer to random noise. The second term explains the variation derived from the imperfect risk classification. More specifically, this improved pure premium, or the refined risk classification based on InsurTech innovation, $\operatorname{E}\left[Y \mid \textbf{X}^{IH}, \textbf{X}^{IT}\right]$, cannot be fully explained by the current rating factors $\textbf{X}^{IH}$ alone. As a result, incorporating additional InsurTech risk factors extends several benefits in insurance ratemaking. Expanded risk factors can help develop a better risk classification and attract more homogeneous policyholders. Hidden risk factors may be uncovered, thereby avoiding premium leakage that can lead to systematic losses. A more comprehensive understanding of risk factors can help the insurer gain relative advantage over its competitors, and further control the variance of the loss.

In addition, we can adopt the credibility theory to further justify the advantages of bringing in external information in the future. \citet{frees2015credibility} address how individual insurers blend external information with their own rating variables to improve the ratemaking process by developing credibility predictions within a generalized linear model framework, with a focus on the Tweedie family, using Bayesian methods.

\section{Model calibration}\label{sec:model}

To date, a wide range of actuarial pricing theories, strategies, and properties have been proposed, ranging from classical statistical methods to modern machine learning techniques. We investigated several algorithms to examine the effectiveness of InsurTech innovation through data enhancement, many of which were unsuitable for our purpose. We meticulously choose to present three representative methodologies related to loss modeling: Tweedie Generalized Linear Model (GLM) in Appendix \ref{subsec::GLM}, Elastic net regularization in Appendix \ref{subsec::GLMnet}, and Light Gradient-Boosting Machine (LightGBM). These three methodologies have been used to find a function $f$ that maps $(\textbf{X}^{IH}, \textbf{X}^{IT}) \stackrel{f}{\longrightarrow} \operatorname{E}\left[Y \mid \textbf{X}^{IH}, \textbf{X}^{IT}\right]$. In other words, supervised learning models predict a response variable, pure premium, from a function, $f$, of risk factors and parameters.

\subsection{Light Gradient-Boosting Machine}\label{subsec::lgbm}

Actuaries nowadays resort to advanced machine learning methods to accurately assess policyholders' risk profile. Machine learning methods provide state-of-the-art performance and high efficiency. \cite{Wuthrich2021} summarize some of these methods for actuarial ratemaking and compare them with traditional actuarial models. We also observe that advanced machine learning methods, including those we considered outside the reach of this paper, can significantly outperform the classical Tweedie GLM using our datasets. 

While machine learning methods are acknowledged for their advanced predictive capabilities, concerns have been mounting regarding their interpretability, especially in light of regulations such as the General Data Protection Regulation (GDPR) in Europe. These regulations emphasize individuals' right to explanation, a crucial consideration in industries like insurance where transparent pricing methods are not only advisable but also mandated by regulators. GBMs, beyond offering competitive predictive performance in our dataset, provide interpretability, offering insights into the model's decision-making process. While each individual tree within a GBM is inherently explainable, the collective result of numerous trees can be intricate. However, as highlighted in \cite{delgado2022implementing}, techniques have been developed to extract feature contributions from GBMs, thereby shedding light on prediction formulations. Such techniques enhance the interpretability of GBMs and align well with regulatory mandates like GDPR, effectively balancing predictive accuracy with model explainability.  With these merits of tree-based models in mind, we now delve into their applications in the actuarial science literature.

\cite{guelman2012gradient} applies GBMs on the prediction of auto at-fault accident losses and demonstrates that it produces a superior predictive accuracy than the traditional GLM approach. \cite{quan2018predictive} model multi-line of business insurance claims data with correlated responses using multivariate tree-based models and show that multivariate tree-based models outperform univariate ones in terms of predictive accuracy and are able to capture the inherent associations among response variables. \cite{lee2018delta} introduce Delta Boosting (DB) and demonstrate that it is the most optimal boosting method for various loss functions and test the proposed DB machine with collision insurance claim data. \cite{yang2018insurance} propose TDBoost, a gradient tree-boosted Tweedie compound Poisson model, and apply the proposed method to the prediction of auto insurance claims. As an extension of TDBoost, \cite{zhou2020tweedie} propose EMTboost, a gradient tree-boosted zero-inflated Tweedie model, for extremely imbalanced data with a large number of zeros. \cite{henckaerts2021boosting} implement tree-based models with tailored loss functions for the development of full tariff plans based on both the frequency and severity of claims and conclude that the boosted trees outperform the classical GLM approach. \cite{hu2022imbalanced} modify the traditional splitting function of decision trees to handle the imbalance problem induced by a large proportion of zero responses in insurance claim datasets and demonstrate the modified tree structure leads to improved prediction accuracy.

More specifically, LightGBM, which is satisfactory for our purpose of seeking an efficient implementation of Gradient Boosting Decision Tree (GBDT), is an ensemble of decision trees (DT) in Appendix \ref{subsec::tree}, where each proceeding tree is grown sequentially based on the negative gradient or residual of previously learned decision trees.

To extend a single DT to GBDT, a series of trees are grown in a sequential fashion, where every tree is targeted to optimize the negative gradient of previous trees. As indicated in \citet{friedman2001greedy}, the true function $f$ can be approximated by 
$$
f^{B}(\textbf{X})=\sum_{b=0}^{B}f_{b}(\textbf{X})
$$ 
where $B$ denotes the number of functions or iterations, $f_{0}$ is the initialization function and $\{f_{b}\}_{b=1}^{B}$ are the incremental functions. In the case of GBDT, each $f_{b}$ is a DT and $f_{b}$ is grown at iteration $b$. In the circumstance of steepest-descent with CART, if we define the sum of trees up until the current step $B$ as $f^{B}(\textbf{X})=\sum_{b=0}^{B}f_{b}(\textbf{X})$, the tree update procedure can be formulated as 
$$
f^{B}(\textbf{X})=f^{B-1}(\textbf{X})+\rho_{B}\sum_{m=1}^{M(B)}\hat{c}_{mB}\mathbbm{1}_{R_{mB}}(\textbf{X})
$$ 
where $\rho_{B}$ is the pre-defined learning rate, $M(B)$ refers to the number of leaf nodes at iteration $B$, $R_{mB}$ is the leaf node $m$ at iteration $B$, $\hat{c}_{mB}$ is the prediction constant of node $R_{mB}$, which is estimated by averaging the node, as stated above.

Furthermore, as stated in \citet{Ke2017}, to reduce the computation complexity induced by the recursive binary splitting process, where each internal node needs to scan every feature and every observation, LightGBM innovatively utilizes Gradient-based One-Side Sampling (GOSS) and Exclusive Feature Bundling (EFB) to address the computation issues. GOSS is a sampling mechanism that weights observations with larger gradients higher, which leads to more contributions to the incremental functions. On the other hand, it employs a down-sampling technique to retain those observations with a smaller gradient, which leads to randomly dropping less influential observations before the tree construction procedure. This method is typically better than random sampling, which assigns the same sampling rate. Furthermore, to reduce the dimension of the feature space, especially when we have a sparse dataset, EFB proposes to combine the exclusive features as bundles whose size is significantly reduced compared to the original feature space. Through GOSS and EFB, \citet{Ke2017} claim to substantially speed up GBDT while maintaining comparable accuracy, which is one of the reasons we select this algorithm given that we have a large dataset with almost 1 million observations and more than 500 features. 

All machine learning methods need proper model calibration, which involves train-test split, cross-validation, and hyperparameter tuning. It requires massive computation and various experiments. We are able to leverage the HAL system\footnote{This work utilizes resources supported by the National Science Foundation’s Major Research Instrumentation program, grant 1725729, as well as the University of Illinois Urbana-Champaign. Retrieved from https://wiki.ncsa.illinois.edu/display/ISL20/HAL+cluster} to perform cloud computing in a secured environment. See \cite{kindratenko2020hal}. In addition, we use the Optuna framework to perform hyperparameter optimization, specifically using Bayesian optimization. See \cite{akiba2019optuna} and \cite{shahriari2015taking}. The Optuna framework is able to fully utilize distributed cloud computing by allowing users to submit multiple batch jobs on the same training process, which narrows down the optimal hyperparameter space with the Bayesian optimization mechanism.

\subsection{Comparison of model performance}
This section compares the performance of these two models: insurance in-house models versus InsurTech-enhanced models. As we mentioned in Subsection \ref{subsec::bop_data}, we model each coverage group separately using LightGBM and Tweedie GLM after elastic net feature selection, and thus we have six InsurTech-enhanced models in total. For LightGBM model calibration, we use Bayesian optimization with the Optuna framework, as discussed in Section \ref{subsec::lgbm}, and the optimization objective loss function is the mean absolute error (MAE). For the Tweedie GLM after elastic net feature selection model calibration, we use grid search to find the optimal hyperparameters mentioned in Appendix \ref{appendix_sec:tweedie}. We apply a 10-fold cross-validation on the training set to find the best models for both methods. For the hyperparameter values of the final models after tuning. See Appendix \ref{appendix_sec:parameter}. 

\begin{figure}[!ht]
\centering
\begin{subfigure}{.48\textwidth}
\centering
\includegraphics[width=0.9\linewidth]{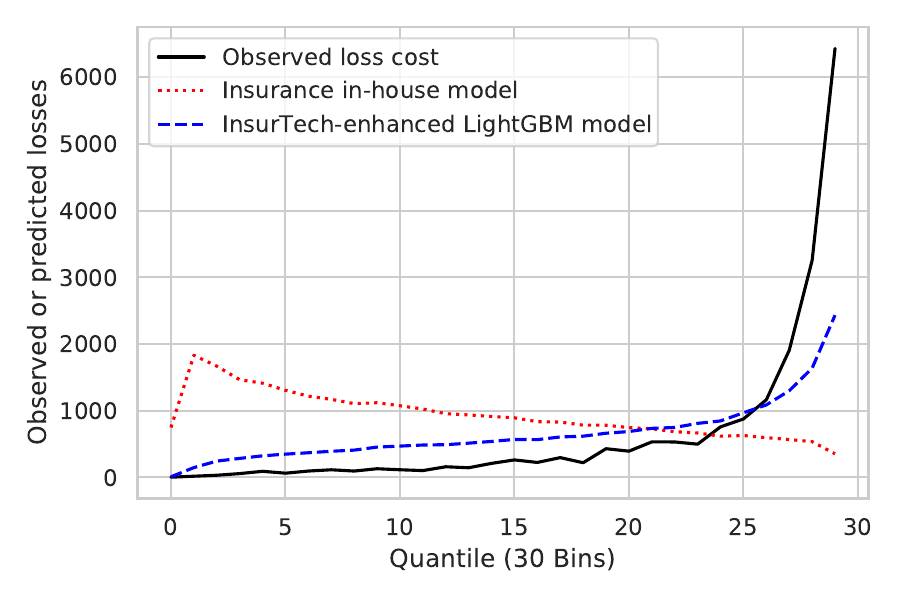}
\caption{Double lift chart for BG train}
\label{fig:double_lift_bg_train}
\end{subfigure}
\begin{subfigure}{.48\textwidth}
\centering
\includegraphics[width= 0.9\linewidth]{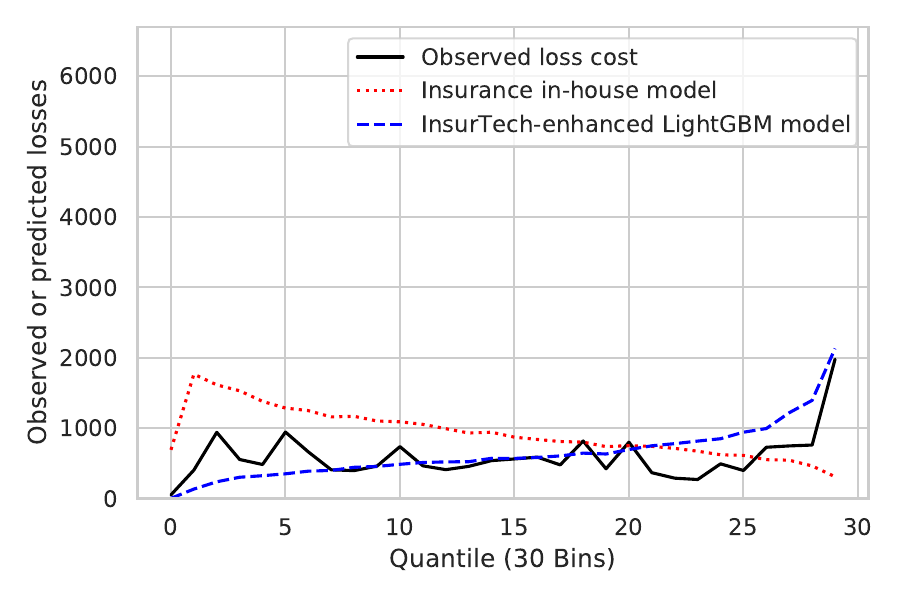}
\caption{Double lift chart for BG test}
\label{fig:double_lift_bg_test}
\end{subfigure}
\begin{subfigure}{.48\textwidth}
\centering
\includegraphics[width=0.9\linewidth]{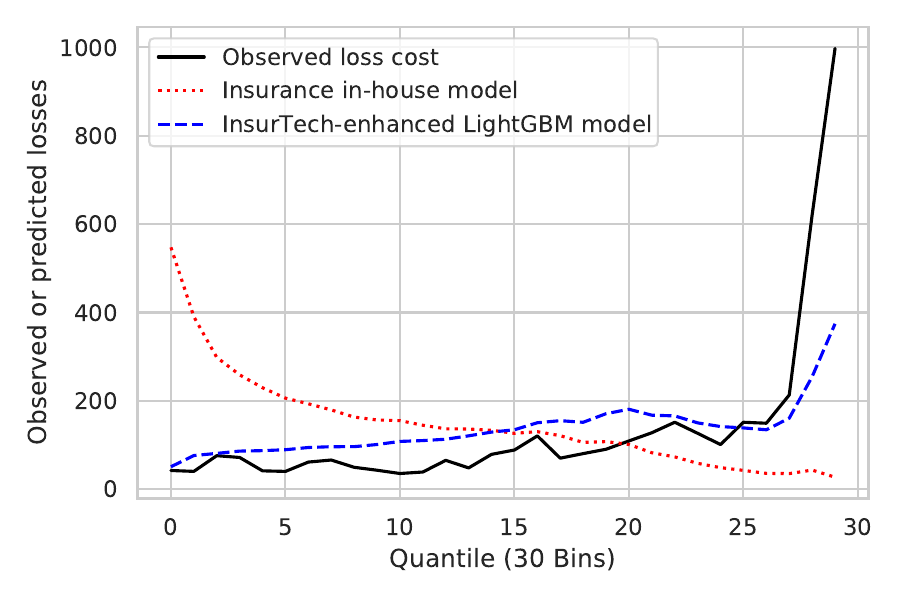}
\caption{Double lift chart for BP train}
\label{fig:double_lift_bp_train}
\end{subfigure}
\begin{subfigure}{.48\textwidth}
\centering
\includegraphics[width= 0.9\linewidth]{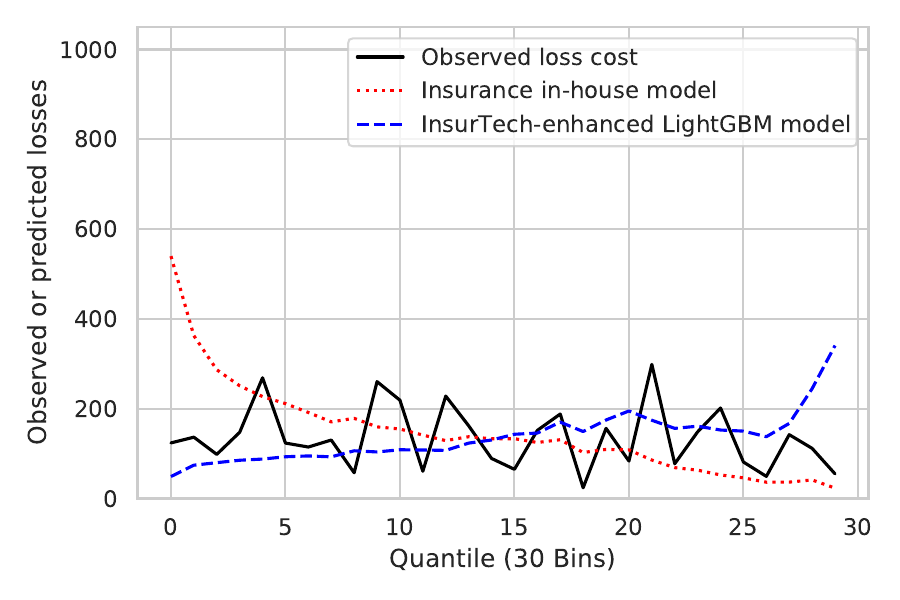}
\caption{Double lift chart for BP test}
\label{fig:double_lift_bp_test}
\end{subfigure}
\begin{subfigure}{.48\textwidth}
\centering
\includegraphics[width=0.9\linewidth]{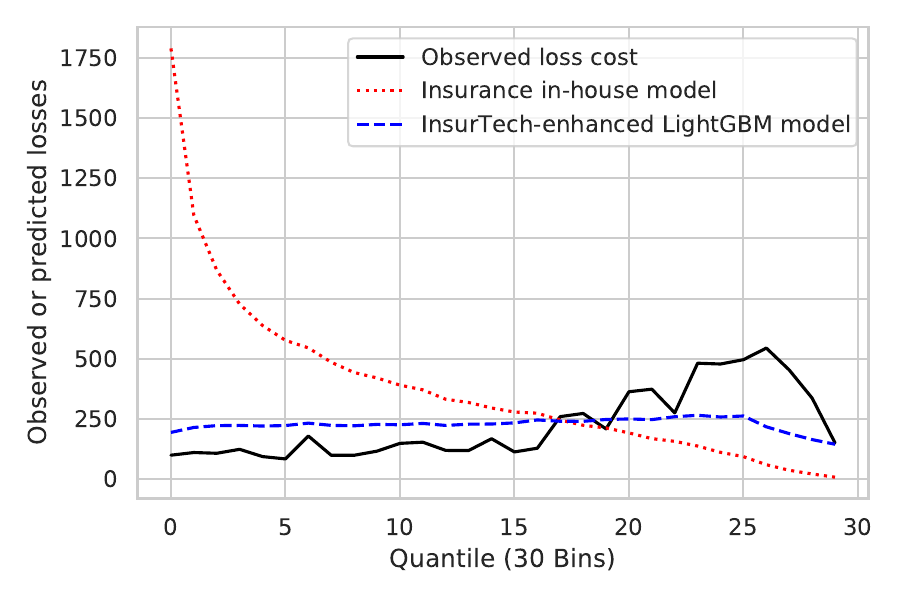}
\caption{Double lift chart for LIAB train}
\label{fig:double_lift_liab_train}
\end{subfigure}
\begin{subfigure}{.48\textwidth}
\centering
\includegraphics[width= 0.9\linewidth]{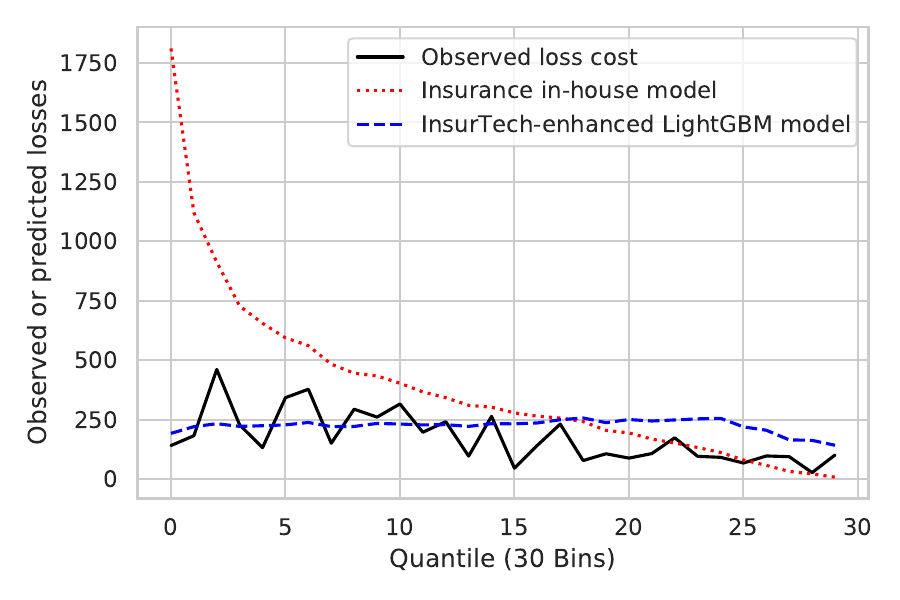}
\caption{Double lift chart for LIAB test}
\label{fig:double_lift_liab_test}
\end{subfigure}
\caption{Double lift charts for model comparison}
\label{fig:double_lift}
\end{figure}

We use double lift charts to visually compare the predictive performance of InsurTech-enhanced models and insurance in-house models. Model lift refers to the ability to differentiate between low- and high-risk policyholders and can be used to measure a model's business value, which helps c-suite leaders make a business decision. Double lift charts are commonly used in insurance companies to measure model lift and compare predictive power between two distinct models. In our experiments, double lift charts are created according to the following procedures: First, we sort the data by a ratio of new model prediction (InsurTech-enhanced model prediction) to the current pure premium (insurance in-house model prediction); Second, the sorted data is subdivided into quantiles with equal exposure (we use 30 quantiles); Lastly, for each quantile, we calculate the average observed loss cost, the average current pure premium, and the average new model predicted loss cost. In addition, we numerically examine predictive accuracy using a variety of validation measures. These measures are listed as follows: Gini index, percentage error (PE), root mean squared error (RMSE), and mean absolute error (MAE). For definitions and interpretations of these validation measures. See Appendix \ref{appendix_sec:VM}. 

Take Subfigure \ref{fig:double_lift_bg_train} and Subfigure \ref{fig:double_lift_bg_test}, the double lift charts of the BG group, for example, the observed loss cost (actual value) is indicated by the solid black line, the predictions provided by insurance in-house models are indicated by the red dotted line, and the predictions derived from InsurTech-enhanced LightGBM models are indicated by the blue dashed line. We can observe that the blue dashed line follows the trend of the solid black line closer, compared to the red dotted line, indicating that InsurTech-enhanced LightGBM models have outperformed insurance in-house models with greatly improved prediction accuracy. Also, it can be seen that InsurTech-enhanced LightGBM models are able to capture the spike of the observed loss cost on the right side, which represents the larger claims. Similarly, from Subfigure \ref{fig:double_lift_liab_train} and Subfigure \ref{fig:double_lift_liab_test}, we can see that the above statement also holds for the LIAB group. For the BP group, the InsurTech-enhanced LightGBM model exhibits a slight edge compared to the insurance in-house model. These empirical results show that the data created by InsurTech innovations can significantly improve the insurance company's loss models and help differentiate the larger claims. While the double lift charts offer a broad perspective on the predictive performance of the model, we explore their exact prediction accuracy in the subsequent discussion by utilizing various validation metrics to gain a more profound understanding of the quantitative improvement of the model.
\begin{table}[!h]
\centering
\begin{tabular}{lllrrrr}
\toprule
Coverage & Dataset & Model & Gini & PE & RMSE & MAE \\
\midrule
\multirow{6}{*}{BG} & \multirow{3}{*}{train} & Insurance in-house model & 0.29 & -0.40 & 5761.94 & 1526.47 \\
& & Tweedie GLM + elastic net & 0.44 & -0.04 & 5660.01 & 1286.31 \\
& & LightGBM & 0.84 & 0.00 & 5364.05 & 1198.07 \\
\cline{2-7}
& \multirow{3}{*}{test} & Insurance in-house model & 0.32 & -0.54 & 5328.02 & 1461.92 \\
& & Tweedie GLM + elastic net & 0.32 & -0.16 & 5284.90 & 1238.94 \\
& & LightGBM & \textbf{0.37} & \textbf{-0.08} & \textbf{5198.57} & \textbf{1181.47} \\
\midrule
\multirow{6}{*}{BP} & \multirow{3}{*}{train} & Insurance in-house model & 0.59 & -0.07 & 2498.13 & 277.37 \\
& & Tweedie GLM + elastic net & 0.68 & 0.00 & 2450.82 & 262.64 \\
& & LightGBM & 0.78 & 0.00 & 2409.88 & 259.11 \\
\cline{2-7}
& \multirow{3}{*}{test} & Insurance in-house model & 0.58 & -0.11 & 2350.80 & 270.75 \\
& & Tweedie GLM + elastic net & 0.36 & \textbf{-0.04} & 2375.10 & \textbf{262.31} \\
& & LightGBM & \textbf{0.59} & -0.06 & \textbf{2348.93} & 262.78 \\
\midrule
\multirow{6}{*}{LIAB} & \multirow{3}{*}{train} & Insurance in-house model & 0.57 & -0.67 & 3937.22 & 586.88 \\
& & Tweedie GLM + elastic net & 0.63 & -0.04 & 3920.13 & 449.25 \\
& & LightGBM & 0.78 & 0.00 & 3853.67 & 435.65 \\
\cline{2-7}
& \multirow{3}{*}{test} & Insurance in-house model & 0.54 & -1.15 & 3347.60 & 547.02 \\
& & Tweedie GLM + elastic net & 0.47 & -0.33 & 3340.86 & 408.15 \\
& & LightGBM & \textbf{0.56} & \textbf{-0.26} & \textbf{3305.85} & \textbf{394.56} \\
\bottomrule
\end{tabular}
\caption{Model performance based on validation measures}
\label{tab:VM-result}
\end{table}

Table \ref{tab:VM-result} compares the predictive performance of InsurTech-enhanced models, Tweedie GLM with elastic net feature selection and LightGBM, and the insurance in-house model based on the training and test datasets of each coverage group. For each coverage, we highlight the best-performing model with the value of each validation measure in bold. In general, both InsurTech-enhanced models consistently outperform the insurance in-house model for each coverage group. This suggests that the improvement stems from the additional information provided by InsurTech, irrespective of the chosen loss model. It should be noted that InsurTech-enhanced models significantly reduce the absolute value of percentage error (PE) of observed loss cost predictions compared to insurance in-house models. This indicates that InsurTech-enhanced models have superior predictive performance at the portfolio level, which is usually a key concern for insurers, particularly from a financial statement perspective.

\section{Model interpretation and discussion}\label{sec:interpretation}
Modern machine learning models have often been perceived as black-box models and are criticized for producing results that may be difficult to interpret, particularly in the actuarial domain. In this section, we demonstrate the actuarial interpretability of our calibrated models by examining feature importance and exploring a selection of illustrative business cases. Although this approach is not without its flaws, we still observe some insightful interpretations that resonate with our expert knowledge. Regarding our naming protocol, features with names preceded by uppercase letters, apart from the initial letter, are drawn from the InsurTech company files; all others have names that only have the initial letter as capitalized and are drawn from the insurance company files.

\subsection{Feature importance}

With more than 500 features in the dataset, we now evaluate how much each feature contributed to the prediction of the InsurTech-enhanced LightGBM model discussed in Section \ref{sec:model}. Three different approaches are applied to assess the importance of features: Mean Decrease in Impurity (MDI), Mean Decrease in Accuracy (MDA), and SHapley Additive exPlanations (SHAP). The MDI, also referred to as Gini feature importance, was detailed in \cite{breiman1984classification}. It measures the feature importance by calculating the average decrease in impurities of all splits in the tree-based model building process. The MDA, also called permutation feature importance, was introduced by \cite{breiman2001random}, which is a model-agnostic global explanation method that measures the importance of a feature by calculating the decrease in a model score (in our case, the increase in the model’s mean squared error) after randomly permuting the given feature with other features. Unlike MDA, \citet{lundberg2017unified} propose SHAP, which measures the local feature importance for every observation by computing Shapley values from coalitional game theory. SHAP is an additive feature attribution method whose model consists of a linear function of simplified binary indication of the entire feature space. The SHAP values are calculated by permuting the existence of each feature and the corresponding model predictions. 

\begin{figure}[h!]
\centering
\begin{subfigure}{0.48\textwidth}
\centering
\includegraphics[width= 0.95\linewidth]{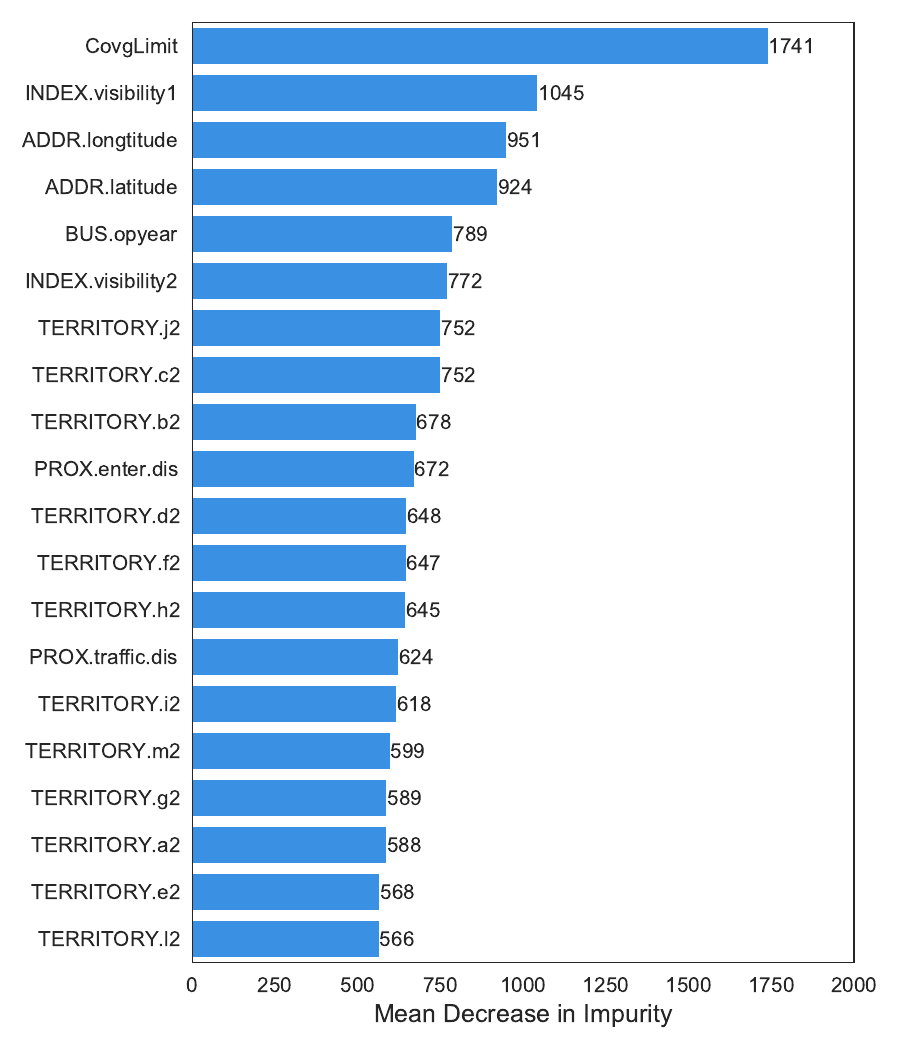}
\caption{Mean Decrease in Impurity - Top 20 features}
\label{fig:gini_fi}
\end{subfigure}
\begin{subfigure}{.48\textwidth}
\centering
\includegraphics[width=0.95\linewidth]{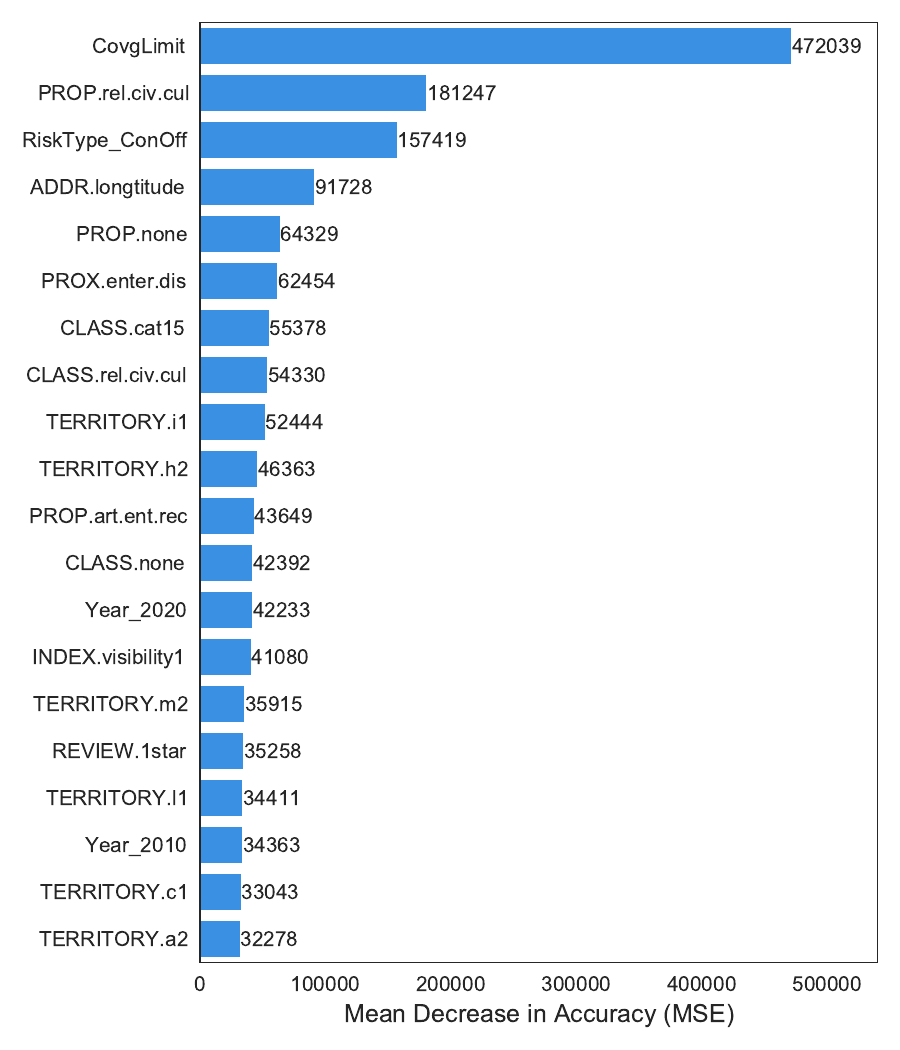}
\caption{Mean Decrease in Accuracy- Top 20 features}
\label{fig:perm_fi}
\end{subfigure}
\begin{subfigure}{.48\textwidth}
\centering
\includegraphics[width= 0.95\linewidth]{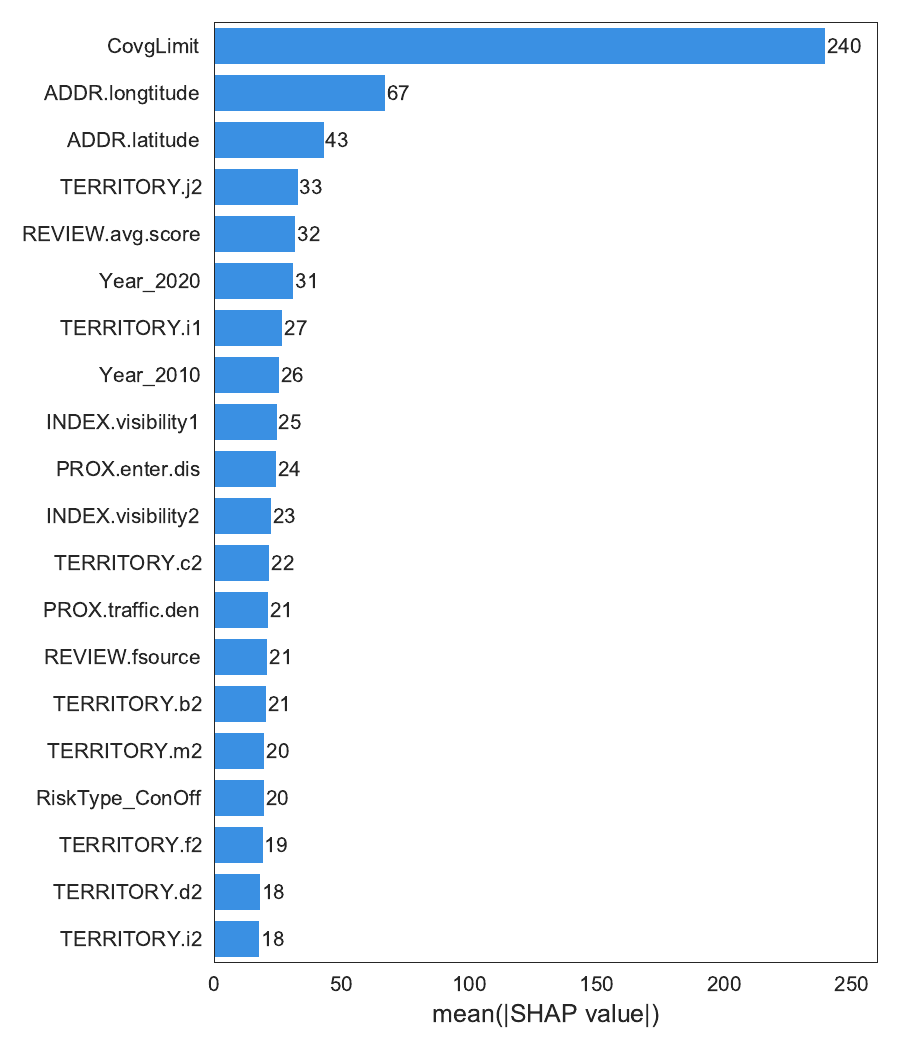}
\caption{SHAP feature importance - Top 20 features}
\label{fig:shap_fi}
\end{subfigure}
\caption{Top 20 influential features in InsurTech-enhanced LightGBM model (BG test)}
\label{fig:feature_importance}
\end{figure}

Figure \ref{fig:feature_importance} highlights the top 20 features calculated by the three methods mentioned above in the InsurTech-enhanced LightGBM model on the BG subdataset. It illustrates the extent of each feature's contribution to the prediction of the model, highlighting the significant features that may reveal potential rating factors that were overlooked in insurance pricing. There is a notable overlap in the top features based on the three different feature importance methods. These top features include the coordinates of addresses (\textit{ADDR.longtitude} and \textit{ADDR.latitude}) in category \textit{Business Information}, proximity scores (i.e., \textit{PROX.enter.dis}, \textit{PROX.traffic.dis}, and \textit{PROX.traffic.den}), and territory risks (i.e., \textit{TERRITORY.xx}), all of which are related to the location of a business; See Table \ref{tab:InsurTech-features}. These location-related risk factors are fairly important to the BG coverage group, as they cover losses related to the building of a business. The territory risks, particularly those in version 2 (e.g., \textit{TERRITORY.j2}, regarding fire risk characteristics), contribute significantly to the model's predictive capability. This is not surprising and is expected, given that territory risk features are engineered from business risk characteristics and thus contain more risk information and have higher fill rates. Other top InsurTech features in the model include visibility indexes (i.e., \textit{INDEX.visibility1} and \textit{INDEX.visibility2}), review scores (i.e., \textit{REVIEW.xx} extracted from text data), business classification (i.e., \textit{CLASS.xx}, the overall business classification assigned to a business), and corresponding segment proportions (i.e., \textit{PROP.xx}, the proportion of total votes for a segment given to a business). Furthermore, the coverage limit (i.e., \textit{CovgLimit}) and risk type (i.e., \textit{RiskType\_xx}) provided by the insurance company are also important rating factors that describe policy information and measure exposure information. Specifically, we can see from Subfigure \ref{fig:perm_fi} and Subfigure \ref{fig:shap_fi} that the feature risk type condo/office (i.e., \textit{RiskType\_ConOff}) is of high importance, which is consistent with our observations in Section \ref{subsec::bop_data} that the BOP policies of condo/office risk type have the highest observed loss cost on average.

In a classical linear model, although it is challenging to avoid pitfalls when multicollinearity and high-dimensional data exist, we are used to interpreting the model using linear coefficients. For black-box models, we have similar tools to observe how one feature has an impact on the response variable by keeping other features unchanged. \citet{apley2020visualizing} introduce Accumulated Local Effects (ALE), which is a global model-agnostic explanation method that evaluates the average impact of a feature on the predictions by integrating the averaged local effect (i.e., the difference in predictions across the conditional distribution). Compared to the widely used Partial Dependence Plots (PDP), ALE is more robust against correlated features and less computationally expensive. Furthermore, there are other techniques to achieve similar goals, such as Individual Conditional Expectation (ICE) plots to display one line per observation.

\begin{figure}[h!]
\centering
\begin{subfigure}{.5\textwidth}
\centering
\includegraphics[width= 0.9\linewidth]{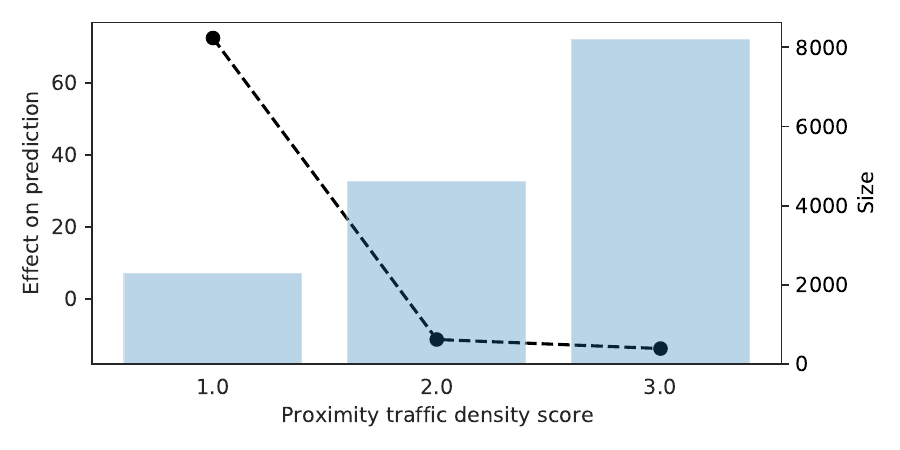}
\caption{Proximity traffic density score}
\label{fig:ale_traffic_density}
\end{subfigure}%
\begin{subfigure}{.5\textwidth}
\centering
\includegraphics[width=0.9\linewidth]{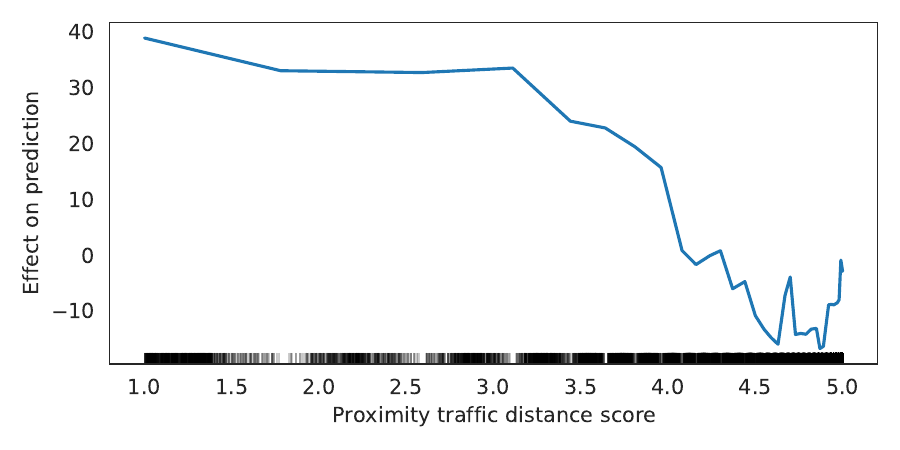}
\caption{Proximity traffic distance score}
\label{fig:ale_traffic_distance}
\end{subfigure}
\caption{ALE plots of proximity traffic scores}
\label{fig:ale_traffic}
\end{figure}

Figure \ref{fig:ale_traffic} presents the ALE plots of proximity traffic scores. The y-axis shows the centered effect on the predictions, which corresponds to the change in the model predictions when the feature is given a particular value relative to the average prediction. Subfigure \ref{fig:ale_traffic_density} shows the ALE plot of the traffic density score, the bar charts represent the size of each category, and the dashed line represents the effect of each value of the density score. With a density score of 1, there is a relatively positive impact on the prediction; however, as the density score increases, the relative effect decreases and becomes negative. Note that this is in line with the definition of the traffic density score, where high scores indicate low risk. Similarly, Subfigure \ref{fig:ale_traffic_distance} presents a downward trend in the effect on predictions as the traffic distance score increases. Furthermore, there is a sharp decrease in the effect of the distance score when it exceeds 3.0, which could serve as an indicator of risk for insurers. This gives us a clearer visualization of how proximity traffic scores impact the observed claim loss compared to the one-way analysis in Figure \ref{fig:traffic}. Specifically, in a one-way analysis, the effect of a particular feature on the response variable is evaluated without considering the effect of other features contained in the dataset. The ALE plot can solve this issue by using model results that take into account correlations in features. 

\subsection{Illustrative individual cases}

\begin{figure}[h!]
\centering
\begin{subfigure}{.48\textwidth}
\centering
\includegraphics[width=0.95\linewidth]{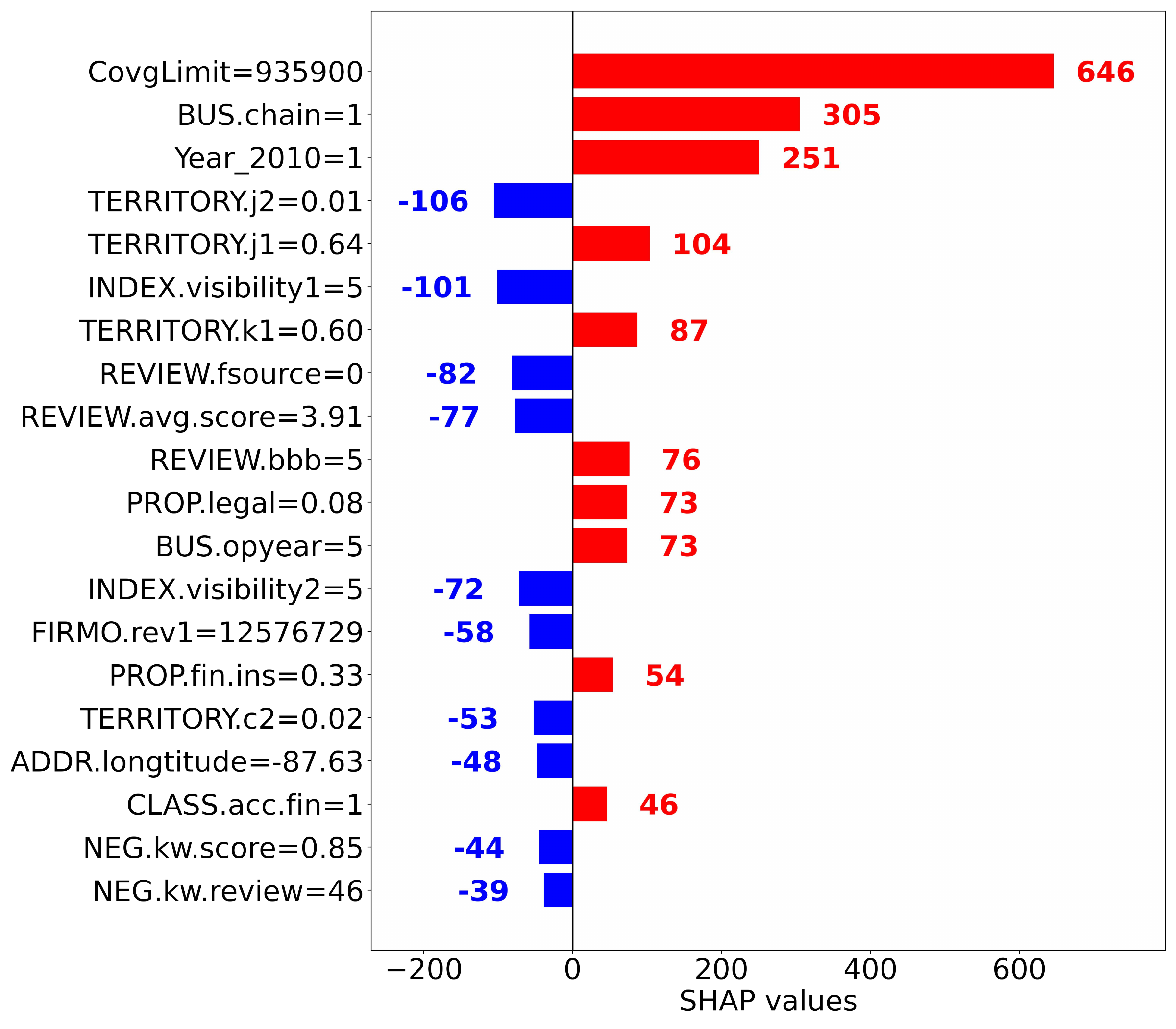}
\caption{Top 20 influential features of Business A}
\label{fig:ind_A_shap}
\end{subfigure}
\begin{subfigure}{.48\textwidth}
\centering
\includegraphics[width= 0.95\linewidth]{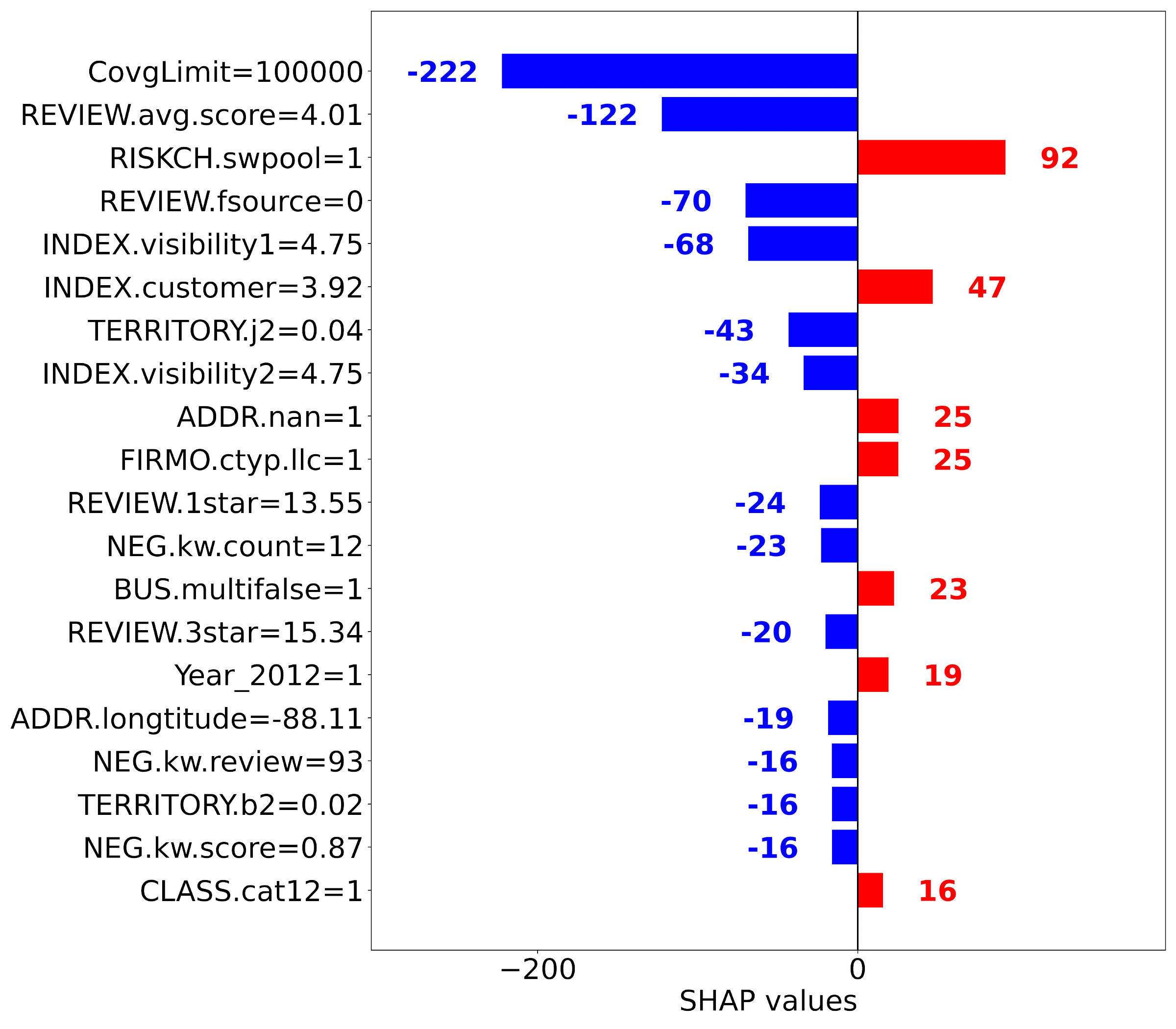}
\caption{Top 20 influential features of Business B}
\label{fig:ind_B_shap}
\end{subfigure}
\begin{subfigure}{.48\textwidth}
\centering
\includegraphics[width=0.95\linewidth]{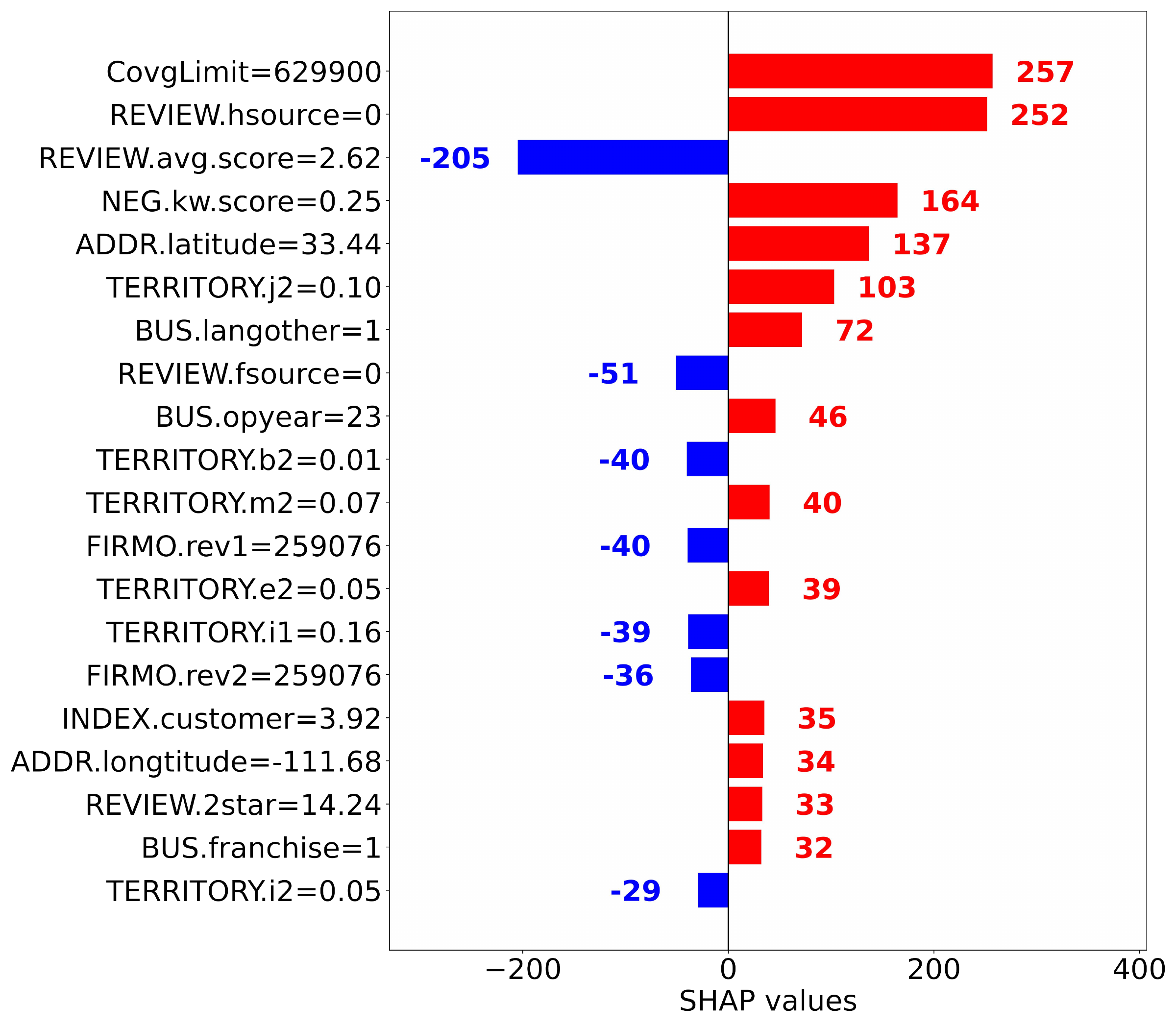}
\caption{Top 20 influential features of Business C}
\label{fig:ind_C_shap}
\end{subfigure}
\begin{subfigure}{.48\textwidth}
\centering
\includegraphics[width= 0.95\linewidth]{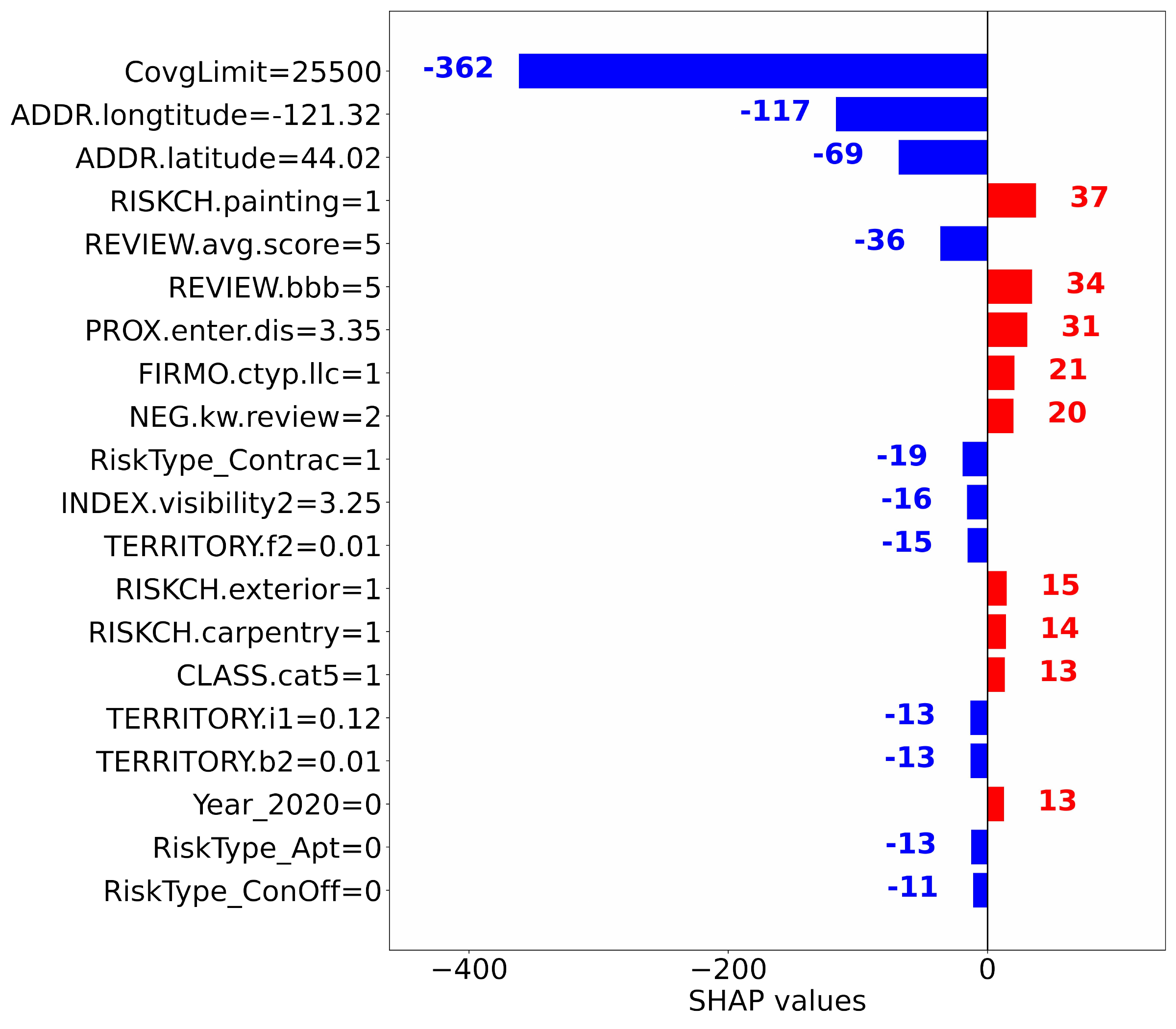}
\caption{Top 20 influential features of Business D}
\label{fig:ind_D_shap}
\end{subfigure}
\caption{Top 20 influential features of 4 local examples}
\label{fig:ind_case}
\end{figure}

In order to examine how the InsurTech risk factors affect the loss model locally, four real businesses are observed from a microscopic point of view. Because it helps enhance model interpretability, we use the SHAP value to explain every individual prediction based on the InsurTech-enhanced LightGBM model. SHAP values quantify the feature importance of all features and explain how each feature impacts individual predictions. Due to space constraints, we only highlight 20 important features among more than 500 features in the context of absolute SHAP values. In Figure \ref{fig:ind_case}, the four real businesses are extracted from our dataset to illustrate how the features influence the observed claim loss. Note that the features labeled as red means positively contribute to the observed claim loss, meaning this feature is relatively riskier than the features labeled as blue. These businesses are described as follows (a) a business with a positive claim from the training dataset; (b) a business with no claim from the training dataset; (c) a business with a positive claim from the test dataset; (d) a business with no claim from the test dataset.

In (a), Business A is a land trust company that provides legal and financial services to real estate or property owners. InsurTech risk factor captures the corresponding business classification and segment information through \textit{CLASS.acc.fin} (business classification is accountants \& financial services), \textit{PROP.legal} (business segment has the proportion of legal services), and \textit{PROP.fin.ins} (the proportion of the finance \& insurance) which are belongs to category \textit{Classification} in Table \ref{tab:InsurTech-features}. Business A has multiple chain businesses statewide, which is indicated by the InsurTech risk factor, \textit{BUS.chain} in category \textit{Business Information}, which may have a strong positive impact on the observed loss cost. In 2010, \textit{Year\_2010}, Business A purchased BOP from our insurance partner, right after five years of operation, \textit{BUS.opyear=5}, which is considered a relatively young company with potential risk. Business A has a corporation revenue of 12576729, based on \textit{FIRMO.rev1} in category \textit{Firmographics}, which means that it has a more significant revenue than average, thus being considered as a less risky business. Index and territory risk further describe the InsurTech risk factor associated with Business A, and these risk factors have a significant impact on determining the observed claim loss from a modeling perspective. We can also see that the \textit{ADDR.longtitude} shows up in the important feature list, since it is correlated with the territory risk, which is defined by zip code information. Similarly, social media and online content provide more detailed information and its contribution to the response variable. For example, Business A has an average review score of 3.91, \textit{REVIEW.avg.score}, Better Business Bureau rating is 5, \textit{REVIEW.bbb} and these reviews exclude source F, \textit{REVIEW.fscource} and all the information comes from the category \textit{Text Data}. InsurTech company also creates negative keyword scores and counts, \textit{NEG.kw.score} and \textit{NEG.kw.review} to summarize these text data using Natural Language Processing techniques.  

Take another example of (b), Business B is a rental apartment. This business joined as a BOP policyholder in year 2012, implied by \textit{YEAR\_2012}, and has not submitted any claims during the period for which the policy is in-force. As advertised on its website, the apartment premises include a swimming pool, a risk captured by \textit{RISKCH.swpool} (business risk characteristics whether the business has a swimming pool) of category \textit{Risk Characteristics}. Industries such as rental apartments are heavily influenced by subjective customer reviews and ratings. Engineered InsurTech features that quantify those scores can help identify the risk. In this example, the overall appearance of Business B across social media, ratings and review sites entails to what extent the business values its targeted Internet advertisement, which is measured by the engineered scores \textit{INDEX.visibility1} and \textit{INDEX.visibility2} of category \textit{Index}. Business B, scored both 4.75, is higher than average, and is considered a safer business in this regard. Similarly, another feature of category \textit{Index}, the customer ratings, \textit{INDEX.customer}, is rated 3.92, a relative score, and is lower than expected from the perspective of a rental apartment, which explains the positive impact on claim assessment. Additionally, detailed score distributions, likes/dislikes (and comments), can boost the accuracy of risk identification. In the common 5-star rating system where 1 star represents the lowest quality of service/products and 5 star suggests the highest, the proportion of each of the star ratings can help distinguish the service provided, and that information is captured by InsurTech features, \textit{REVIEW.1star} through \textit{REVIEW.5star} of broader category \textit{Text Data}. A 13.55 score of \textit{REVIEW.1star}, which is lower than average, and a 15.34 score of \textit{REVIEW.3star}, which is higher than expected, are consistent with a high average (4.01) score of \textit{REVIEW.ave.score} summarized from previous customers' review contents for Business B. Overall, it has a safer business operation.

The claim submitted by Business C demonstrated in example (c) further strengthens the point that the insurance industry can benefit from InsurTech features for insurance risk assessment. Business C is a licensed medical clinic specializing in Internal Medicine, which purchased a BOP policy to insure its business operations against potential risks. The insured clinic is a part of a franchise business, indicated by \textit{BUS.franchise}, suggesting probable operation risk. The service business provided is also highly correlated with risk evaluation. As one of those identifying criteria, \textit{BUS.langother} is a feature that distinguishes whether the business supports less commonly used languages in the United States. In the example of Business C, this implies another complication in medical practice and thus a more risky business operation. Similarly to previous examples, internet content plays a significant role for the business. Today, prospective patients have a habit of relying on internet reviews to help them decide which physicians to visit. The score, \textit{REVIEW.hsource} of class \textit{Text Data}, denotes the number of scores from Source H, which is an online information platform specialized for physicians and hospitals. Counted 0 from Business C indicates that fewer customer reviews appear than expected and, correspondingly, a significant positive impact on claim prediction.

Belonging to one of the major business lines, Contractors, Business D in (d) is a painting company that, as advertised, provides interior / exterior painting and color design services. The business, as its legal status stands, is a Limited Liability Company (LLC), suggested by \textit{FIRMO.ctyp.llc}, which differentiates whether the company is an LLC. The limited liability property suggests potential risk, which contributes positively to loss prediction. The features, \textit{RISKCH.painting}, \textit{RISKCH.exterior}, and \textit{RISKCH.carpentry} from the same category \textit{Risk Characteristics}, denote the identification of risk characteristics whether the business provides painting, exterior design, and carpentry services, respectively, and accurately capture the characteristics of the business. The score in category \textit{Proximity Score}, \textit{PROX.enter.dis}, is an engineered feature quantifying nearby entertainment-related risk measured by distance. Business D has a higher risk related to entertainment than expected, while other location-based risk is lower and therefore considered a safe business. 

Across all 4 examples, some of the other InsurTech features reveal a consistent impact on predictions. Both Business B and D are categorized as Limited Liability Company (LLC), and both contribute positively to loss predictions. Visibility scores by features \textit{INDEX.visibility1}, \textit{INDEX.visibility2} and customer ratings \textit{INDEX.customer}, can be found in all 4 examples, and a higher overall appearance of the business on the Internet (higher visibility score), or higher customer feedback can be interpreted as lower business operation risk, and thus a negative influence on final predictions. In the common 5-star rating systems where 5 star stands for the highest quality of products/services represented by the features \textit{REVIEW.1star} through \textit{REVIEW.5star}, as seen from Business B and C, a higher proportion of low stars (1 star or 2 stars) implies less trust from customers and thus a higher possibility of operation risk. Conversely, a higher proportion of positive reviews suggests a safer business operation.

Apart from InsurTech features, features such as coverage limit \textit{CovgLimit} and groups/risk types provided by insurance companies and used as rating factors still play a significant role in each of the interpretations, as they are comparatively important in describing policy content. However, to better characterize individual policies, InsurTech features outperform many of the traditional insurance features that the insurance company uses as rating factors in our empirical study.

\section{Concluding remarks}\label{sec:conclude}

The continuing growth in InsurTech innovations is transforming the insurance industry in its manner of embracing rapid advances in technology to modernize the entire insurance value chain. Such modernization results in the flow of big data that allows InsurTech to offer innovative data-driven solutions across all major lines of business in the insurance industry. This paper illustrates how to leverage InsurTech innovations to enhance the forecasting of insurance loss costs and quantifies the improvements from such innovations. Using a portfolio comprising nearly a million BOP policies, this work not only proposes advancements in InsurTech-enhanced loss modeling, but also advocates for the integrating benefits derived from both traditional rating factors used by an insurance company and external risk factors collected by an InsurTech company. A critical challenge is combining and centralizing these data sources, and through a successful academic-industry collaboration of the University of Illinois Urbana-Champaign, Carpe Data (an InsurTech company), and an anonymous insurance company, such work is made possible.

An InsurTech-enhanced loss model plays a pivotal role in an insurance company's quest to reduce the loss ratio, and advancements in this model, for which this paper proposes, significantly help a company manage risks more effectively. Its capacity to accurately assess risk enables insurers to evaluate associated risks of diverse policies and policyholders with greater precision. Such precision extends to premium determination, ensuring policyholders are charged amounts closely aligned with expected losses thereby minimizing the risk of underpricing. Early identification of potential claims allows insurers to proactively mitigate risks, implement preventive measures, or adjust premiums. Furthermore, the adaptability of an InsurTech-enhanced loss model to changing market conditions and emerging risks enables timely adjustments to underwriting and pricing strategies, which can be crucial for staying ahead of evolving risks, preventing adverse selection, and maintaining a balanced portfolio for a lower loss ratio over time. In summary, an enhanced loss model improves an insurance company's ability to assess risk accurately, set appropriate premiums, and proactively manage potential losses, ultimately leading to a more favorable loss ratio essential for financial health and profitability.

We envision that insurance companies may face challenges when adopting the work proposed here to seek regulatory approval for premium rating and risk classification. However, such challenges can be readily overcome with the proper justification. Model interpretability revealed that many relevant and intuitive features that affect loss costs are derived from the InsurTech dataset, including firmographics, risk characteristics, indexes, proximity, and territory risk scores. In addition, using visualization, feature importance, and illustrative individual cases may help improve the understanding of modern machine learning algorithms.

\section*{Acknowledgment}
The authors would like to thank the anonymous insurance company for sharing the BOP insurance data, Carpe Data for providing InsurTech-empowered risk information, and the National Center for Supercomputing Applications for offering computing resources and technical support. In particular, the authors thank Lois, Dan, Steven, Michael, Eli, Perri, Eric, Shuijing, Derek, Mingxi, and Volodymyr Kindratenko (NCSA) for their expertise and assistance. Please note that the last names of the above contributors are omitted due to privacy issues.

\clearpage

\bibliographystyle{apalike}
\bibliography{ref.bib}

\clearpage

\begin{appendices}
\section{Traditional loss models}\label{appendix_sec:tweedie}

\subsection{Tweedie Generalized Linear Model}\label{subsec::GLM}

Generalized Linear Models (GLM), proposed by \citet{nelder1972generalized}, introduce a linear or non-linear relationship between features and response variables through a link function and a distribution function. GLM is a widely accepted actuarial practice because of its advantages in implementation and interpretation, both of which are motivated by the fact that insurance is a strictly regulated industry. The traditional practice of actuarial ratemaking is to find proper loss distributions under the GLM framework. Since the distribution of insurance loss typically contains a point mass at zero representing no claims and a continuous component (often substantial) for positive values representing the amount of a claim, Tweedie GLM has been one of the most frequently used actuarial ratemaking models in the past few decades. See \cite{jorgensen1994fitting}.

With a general loss model framework,
$$
Y=\sum_{i=1}^{N}Z_{i}
$$
where $N$ refers to the number of claims and $Y$ denotes the total claim amount, so that $Z_{i}$ captures the claim severity of each claim, Tweedie distribution, which is categorized as a class of compound Poisson-Gamma distribution, is a special case of such general loss model where $N$ is Poisson distributed and $Z_{i}$ ($i=1,2,...,N$) are independently identically distributed Gamma random variables. Tweedie distribution is a subclass of the family of Exponential Dispersion (ED) models, which can be written as 
$$
f(y \mid \mu, \tau, \phi)=a(y, \phi) \exp \left(\frac{1}{\phi}\left(\frac{y \mu^{1-\tau}}{1-\tau}-\frac{\mu^{2-\tau}}{2-\tau}\right)\right)
$$
where $\phi$ is the dispersion parameter in $(0,+\infty)$, mean $E(Y)=\mu$ and variance $Var(Y)=\phi\mu^{\tau}$, $\tau$ is an extra parameter that controls the variance of the distribution. In the special case of $p=0$, the distribution degenerates to the normal distribution; when $p=1$, it becomes a Poisson distribution; when $p=2$, it is equivalent to Gamma distribution; when $p=3$, it is an inverse Gaussian distribution.

Here, we assume that the dataset consists of a vector of $p$ features, $x_{1}, x_{2}, ..., x_{p}$, and a response variable $y$; that is, our data can be represented as
$$
\left(\textbf{X},Y\right) = \left((\textbf{X}^{IH}, \textbf{X}^{IT}),Y\right)= \{\textbf{x}_{i}, y_{i}\}_{i} = \{(x_{i 1}, x_{i 2} \ldots, x_{i p}), y_i\}_i,
$$
where $i=1, \dots, n$ and $n$ is the number of observations. Once the choice of distribution and link function (logarithmic link) is identified, the regression coefficients, $\beta_{i}$ ($i=1,2,...,p$), can be found by the maximum likelihood estimation. Specifically, under the Tweedie GLM framework, the negative log-likelihood can be written as
\begin{equation}
l\left(\boldsymbol{\beta}\right)=\sum_{i=1}^n w_i\left(\frac{y_i e^{-(\tau-1)\sum_{i=1}^{p}x_{i}\beta_{i}}}{\tau-1}+\frac{e^{(2-\tau)\sum_{i=1}^{p}x_{i}\beta_{i}}}{2-\tau}\right)
\label{eq:negloglikelihood}
\end{equation}
where $w_i$ is the weight for each observation (by default, equal weight is assigned); here, it assumes that $\phi$ is the same for all observations, which is questionable in our real-life dataset. Furthermore, we use our dataset to estimate the mean by $log(\mu)=\sum_{i=1}^{p}x_{i}\beta_{i}$.

However, given the characteristics of our real-life dataset, the Tweedie GLM is not an ideal model, as we have over 500 risk factors requiring an overwhelming feature selection task, and there is also a problem of missing data, which is almost inevitable, demanding extensive imputation tasks.

\subsection{Tweedie Generalized Linear Model with elastic net}\label{subsec::GLMnet}

\cite{qian2016tweedie} examine the group elastic net regularization for Tweedie GLM and proposes an efficient algorithm to model high-dimensional datasets. In our experiments, we also utilize elastic net regularization to reduce the dimension of the feature space. Initially proposed by \citet{Zou2005}, the elastic net has served as a variable selection method through a combination of L1 and L2 regularization on regression coefficients. Under the same setting as in Appendix \ref{subsec::GLM}, the elastic net minimizes the objective function
$$
\boldsymbol{\beta}=\operatorname{argmin}_{\boldsymbol{\beta}} l\left(\boldsymbol{\beta}\right)+\lambda[(1-\alpha)\dfrac{1}{2}\left\|\boldsymbol{\beta}\right\|_{\ell_{2}}^{2}+\alpha\left\|\boldsymbol{\beta}\right\|_{\ell_{1}}]
$$
where $l\left(\boldsymbol{\beta}\right)$ is defined as Equation (\ref{eq:negloglikelihood}), with L1 norm $\left\|\beta\right\|_{\ell_{1}}$ and L2 norm $\left\|\beta\right\|_{\ell_{2}}$ of the regression coefficients, $\lambda \ge 0$ and $ 0 \le \alpha \le 1$ are tuning parameters. If $\alpha=1$, the optimization reduces to Lasso regression and if $\alpha=0$, the optimization reduces to Ridge regression. 

After standardizing all the features, we assume that the more influential features have larger coefficients in absolute values. To achieve the goal of feature selection, variables whose coefficients generated by elastic net larger than some predefined threshold $\beta^{thres}$ are selected for further modeling. After feature selection, we use the subset of our dataset to build the Tweedie GLM. This modeling strategy has two benefits: first, we preserve the original feature scale when we build the Tweedie GLM, and the coefficients are more interpretable compared to directly using elastic net modeling; second, the Tweedie GLM is one of the current ratemaking tools approved by regulators.

\subsection{Decision trees: Classification and Regression Tree (CART)}\label{subsec::tree}

We follow the notation used in \citet{hastie2009elements}. The base learner, DT, is an algorithm that partitions data into $M$ regions through the feature space recurrently by layers of binary internal nodes into disjoint leaf nodes $R_{1}, ..., R_{M}$, where each leaf node $R_{m}$ is assigned a constant $c_{m}$ ($m=1, 2, ..., M$) as prediction. This indicates that a dataset $\left(\textbf{X},Y\right)$ can be modeled by
$$
f(\textbf{x}_{i})=\sum_{m=1}^{M}c_{m}\mathbbm{1}(\textbf{x}_{i}\in R_{m})
$$ 
where $\mathbbm{1}$ is an indicator function and $\mathbbm{1}(\textbf{x}_{i}\in R_{m})$ is equal to 1 when $\textbf{x}_{i}\in R_{m}$. Each leaf node $R_{m}$ consists of the conjunction of decision regions of the preceding internal nodes and the constant $c_{m}$ of prediction that minimizes node impurity. The structure of internal nodes and leaf nodes is grown as the result of recursive binary splitting and stopping criterion. See \cite{Loh2011}. 

Classification and Regression Tree (CART) \citep{breiman1984classification} in the case of regression, optimizes the sum of squared loss 
$$
\sum_{i: \textbf{x}_{i}\in R_{m}}(y_{i}-f(\textbf{x}_{i}))^2
$$ 
as the impurity of node $R_{m}$, for which solves for the estimation of constant $c_{m}$ in the form 
$$
\hat{c}_{m}=\dfrac{1}{|R_{m}|}\sum_{i: \textbf{x}_{i}\in R_{m}}y_{i}=ave(y_{i}|\textbf{x}_{i}\in R_{m})
$$ 
where $|R_{m}|$ refers to the cardinality of the node $R_{m}$ and $ave(\cdot)$ denotes the average. The splitting procedure is designed to gain the most information possible throughout the split. The split at the initial node is to find such $j$ ($j=1,2,...,p$) indicating feature $x_{\cdot j}$  and decision boundary $s$ so that in the disjoint split regions $R_{left}=\{\textbf{x}_{i}| x_{ij}\leq s\}$ and $R_{right}=\{\textbf{x}_{i}| x_{ij}>s\}$, the objective function, controlled by $j$ and $s$, is minimized. The objective function can be expressed as 
$$
\operatorname*{arg\,min}_{j, s} \sum_{i: \textbf{x}_{i}\in R_{left}}(y_{i}-\hat{c}_{R_{left}})^2+\sum_{i: \textbf{x}_{i}\in R_{right}}(y_{i}-\hat{c}_{R_{right}})^2
$$ 
where $\hat{c}_{R_{left}}=\dfrac{1}{|R_{left}|}\sum_{i: \textbf{x}_{i}\in R_{left}}y_{i}$, $\hat{c}_{R_{right}}=\dfrac{1}{|R_{right}|}\sum_{i: \textbf{x}_{i}\in R_{right}}y_{i}$, respectively. For the following internal nodes, the same splitting procedure is applied recursively to the sub-regions split by preceding decision rules to find the optimal feature and the decision boundary within the sub-regions. The tree is then grown after each successful split and gained information represented by the global loss reduction. As indicated by \citet{Song2015}, to prevent over-complex tree structure and over-fitting, and to maintain time efficiency, stopping rules like the minimum number of observations in a leaf node, and the depth of a tree are practical ones to regularize individual tree sizes.

\section{Model Hyperparameter Setting}\label{appendix_sec:parameter}
\begin{table}[!ht]
\centering
\begin{tabular}{l l rrrr rrrr rrrr} \hline
Model & Hyperparameter & Value (BG) & Value (BP) & Value (LIAB)
\\ [0.5ex]
\hline
\multirow{4}*{Tweedie GLM + elastic net }& alpha & 0.0249 & 0.1016 & 1.2602 \\[-0.1ex]
& l1\_ratio & 0.0860 & 0.2228 & 0.3844 \\[-0.1ex]
& coef\_threshold & 1.0 & 1.0 & 0.1 \\[-0.1ex]
& $p$ & 1.1341 & 1.0004 & 1.3357 \\[-0.1ex]\cline{2-5}
\multirow{10}*{LightGBM}& feature\_fraction & 0.5944 & 0.4482 & 0.4177 \\[-0.1ex]
& learning\_rate & 0.0114 & 0.0455 & 0.0023 \\[-0.1ex]
& max\_depth & 28 & 29 & 24 \\[-0.1ex]
& min\_child\_samples & 80 & 80 & 150 \\[-0.1ex]
& num\_leaves & 120 & 40 & 50 \\[-0.1ex]
& reg\_alpha & 0.4056 & 0.0892 & 0.7631 \\[-0.1ex]
& reg\_lambda & 0.9800 & 0.9626 & 0.8913 \\[-0.1ex]
& subsample & 0.5192 & 0.5215 & 0.9319 \\[-0.1ex]
& subsample\_for\_bin & 120000 & 100000 & 60000 \\[-0.1ex]
& n\_estimators & 181 & 67 & 999 \\[-0.1ex]\hline
\end{tabular}
\caption{Model Hyperparameter Setting}
\label{tab: hyperparameter}
\end{table}

Explanation of hyperparameters optimized in the Tweedie GLM after elastic net feature selection

\begin{enumerate}
    \item alpha: alpha is the constant that controls penalization by multiplying both the L1 norm and the L2 norm regularization term.
    \item l1\_ratio: l1\_ratio is proportion of L1 norm penalization. Correspondingly, L2 norm is regularized by $1-\textrm{l1\_ratio}$. If $\textrm{l1\_ratio}=0$, elastic net becomes equivalent to the Ridge regression; if $\textrm{l1\_ratio}=1$, elastic net reduces to the Lasso regression; if $0<\textrm{l1\_ratio}<1$, it is a mixture of L1 and L2 regularization.
    \item coef\_threshold: coef\_threshold is the threshold of feature selection. Variables whose absolute values of coefficients are larger than this threshold are selected for the Tweedie GLM.
    \item $p$: $p$ refers to the variance power that decides the variance of Tweedie distribution. 
\end{enumerate}

Explanation of hyperparameters optimized in LightGBM\footnote{Full reference of hyperparameters by LightGBM. Retrieved from https://lightgbm.readthedocs.io/en/latest/Parameters.html/}:

\begin{enumerate}
    \item feature\_fraction: feature\_fraction is a constant of range $0<\textrm{feature\_fraction}\leq 1$ referring to the proportion of features used for each tree growing. When $\textrm{feature\_fraction}=1$, all features are used.
    \item learning\_rate: learning\_rate is a positive real number that controls the learning step of each tree update.
    \item max\_depth: max\_depth denotes the maximum number of layers of each tree. This hyperparameter controls the size of trees to reduce training time and prevent over-fitting.
    \item min\_child\_samples: min\_child\_samples denotes the minimum number of samples needed for further node split. The hyperparameter is also critical to control the size of trees.
    \item num\_leaves: num\_leaves indicates the maximum number of leaf nodes allowed in a tree. This hyperparameter limits the complexity of each tree in the ensemble.
    \item reg\_alpha: reg\_alpha denotes the scale of L1 regularization.
    \item reg\_lambda: reg\_lambda refers the parameter of L2 penalization.
    \item subsample: subsample is the hyperparameter referring to bagging fraction where only a proportion of data is sampled for each tree split to speed up the training process.
    \item subsample\_for\_bin: subsample\_for\_bin is the number of samples used to construct discrete feature bins.
    \item n\_estimators: n\_estimators implies the number of trees or iterations in the boosting ensemble. More trees grown represent more complex the boosting ensemble.
\end{enumerate}

\section{Validation Measures}\label{appendix_sec:VM}
\begin{itemize}
    \item \textbf{Gini Index.} Gini $= 1 - \dfrac{2}{N-1} \left(N - \dfrac{\sum_{i=1}^N \; i\tilde{y}_i}{\sum_{i=1}^N \tilde{y}_i}\right)$, where $\tilde{y}$ represents the corresponding observed value $y$ based on the ranking of the predicted values $\widehat{y}$. The higher the Gini, the better.
    \item \textbf{Percentage Error.} PE $= \dfrac{\sum_{i=1}^N \widehat{y}_i-\sum_{i=1}^N y_i}{\sum_{i=1}^N y_i}$. The lower the |PE|, the better.
    \item \textbf{Root Mean Squared Error.} RMSE $= \sqrt{\dfrac{1}{N} \sum_{i=1}^N  (\widehat{y}_{i}-{y}_{i})^2 }$. The lower the RMSE, the better.
    \item \textbf{Mean Absolute Error.} MAE $= \dfrac{1}{N} \sum_{i=1}^N  \left| \widehat{y}_{i}-{y}_{i} \right|$. The lower the MAE, the better.
\end{itemize}

\end{appendices}

\end{document}